\documentclass[10pt,letterpaper]{article}
\usepackage[top=0.85in,left=2.75in,footskip=0.75in]{geometry}
\usepackage{amsmath,amssymb}
\usepackage{changepage}
\usepackage[utf8x]{inputenc}
\usepackage{textcomp,marvosym}
\usepackage{cite}
\usepackage{nameref,hyperref}
\hypersetup{
    colorlinks=false,
    pdfborder={0 0 0},
    pdfborderstyle={/S/U/W 0},
}
\usepackage[right]{lineno}
\usepackage{microtype}
\DisableLigatures[f]{encoding = *, family = * }
\usepackage[table]{xcolor}
\usepackage{array}
\newcolumntype{+}{!{\vrule width 2pt}}
\newlength\savedwidth

\usepackage{setspace} 
\raggedright
\usepackage[aboveskip=1pt,labelfont=bf,labelsep=period,justification=raggedright,singlelinecheck=off]{caption}

\makeatletter
\renewcommand{\@biblabel}[1]{\quad#1.}
\makeatother
\usepackage{lastpage,fancyhdr,graphicx}
\usepackage{epstopdf}
\fancyheadoffset[L]{2.25in}
\fancyfootoffset[L]{2.25in}

\usepackage{bm}
\newcommand{\dd}{\mathrm{d}}
\newcommand{\T}{\mathcal{T}}

\begin{document}
\vspace*{0.2in}

\begin{flushleft}
{\Large
\textbf\newline{Integral, mean and covariance of the simplex-truncated multivariate normal distribution}
}
\newline
\\
Matthew P. Adams\textsuperscript{1,2,3*}
\\
\bigskip
\textbf{1} School of Mathematical Sciences, Queensland University of Technology, Brisbane, Queensland, Australia
\\
\textbf{2} Centre for Data Science, Queensland University of Technology, Brisbane, Queensland, Australia
\\
\textbf{3} School of Chemical Engineering, The University of Queensland, St Lucia, Queensland, Australia
\\
\bigskip
* mp.adams@qut.edu.au

\end{flushleft}

\clearpage

\section*{Abstract}
Compositional data, which is data consisting of fractions or probabilities, is common in many fields including ecology, economics, physical science and political science. If these data would otherwise be normally distributed, their spread can be conveniently represented by a multivariate normal distribution truncated to the non-negative space under a unit simplex. Here this distribution is called the simplex-truncated multivariate normal distribution. For calculations on truncated distributions, it is often useful to obtain rapid estimates of their integral, mean and covariance; these quantities characterising the truncated distribution will generally possess different values to the corresponding non-truncated distribution. In this paper, three different approaches that can estimate the integral, mean and covariance of any simplex-truncated multivariate normal distribution are described and compared. These three approaches are (1)~naive rejection sampling, (2) a method described by Gessner \textit{et al}.\ that unifies subset simulation and the Holmes-Diaconis-Ross algorithm with an analytical version of elliptical slice sampling, and (3) a semi-analytical method that expresses the integral, mean and covariance in terms of integrals of hyperrectangularly-truncated multivariate normal distributions, the latter of which are readily computed in modern mathematical and statistical packages. Strong agreement is demonstrated between all three approaches, but the most computationally efficient approach depends strongly both on implementation details and the dimension of the simplex-truncated multivariate normal distribution. For computations in low-dimensional distributions, the semi-analytical method is fast and thus should be considered. As the dimension increases, the Gessner \textit{et al}.\ method becomes the only practically efficient approach of the methods tested here.


\section*{Introduction}
Multivariate normal distributions (MNDs) that are truncated on some spatial domain, have many applications including usage in nonlinear generalised mixed models~\cite{Lee2020}, Bayesian inference~\cite{Robert1995}, and more generally across a range of data analysis problems (e.g.~\cite{Nath1972,Muthen1990,Serra2020}). The unit simplex is a convex polytope of interest for truncating MNDs, because parameters on this polytope can be interpreted as fractions or probabilities~\cite{Altmann2014}, and so can represent compositional data in many fields, including ecology, economics, physical science and political science~\cite{Tsagris2018}. On a unit simplex, all parameters must be greater than or equal to zero, and their sum must be equal to one. This matches with the interpretation of compositional data being composed of non-negative fractions or probabilities that all must sum to one.

In addition to MNDs truncated on a unit simplex, other probability distributions which can model compositional data include the Dirichlet~\cite{Douma2019} and logistic-normal distributions~\cite{Aitchison1980}. These distributions are already bounded by a unit simplex, and may be unimodal or multimodal depending on their parameterisation. However, for some applications it may be necessary to guarantee that the probability distribution is unimodal for all possible parameter values: an MND bounded by a unit simplex possesses this property. The Dirichlet and logistic-normal distributions also cannot deal with zero and one values~\cite{MartinFernandez2011}, which is not an issue for random sampling but may become a problem when real data containing zero or one values are being compared to these distributions. This problem has motivated the development of modified Dirichlet and logistic-normal distributions that can deal with zero or one values \cite{Butler2008,Tsagris2018}. On the other hand, a MND truncated on a unit simplex requires no modifications to deal with zero or one values. The defining parameters of a MND truncated on a unit simplex -- the mean and covariance of the corresponding non-truncated MND -- are also easily interpretable. However, unlike the Dirichlet and logistic-normal distributions, the exact density of an MND truncated on a unit simplex cannot be rapidly calculated, and unlike the Dirichlet distribution, the mean and covariance of an MND truncated on a unit simplex is difficult to calculate. Hence, compared to other similar probability distributions, there are additional computational challenges and costs associated with using an MND truncated on an unit simplex, that need to be overcome to utilise the potential advantages of this distribution.

Any MND truncated on a unit simplex can always be conveniently simplified via linear transformation, without any loss of information, to a MND possessing one less dimension and truncated to the non-negative space \textit{under} a simplex of one less dimension~\cite{Altmann2014}. For compositional data, the existence of this transformation is recognition of the fact that one of the fractions or probabilities provides redundant information, because any of the fractions or probabilities is always equal to one minus the sum of all other fractions or probabilities. Such a transformation also has additional convenience when one of the compositional parameters is of less interest, for example if the chosen redundant parameter represents the volume fraction of a liquid mixture occupied by water, or if it represents the unvegetated fraction of land cover occupied by multiple vegetation types. After the transformation, all remaining parameters still follow a truncated MND and must be greater than or equal to zero, but their sum must now be equal to \textit{or less than} one. In this paper MNDs truncated on these non-negative domains under a unit simplex are examined; these distributions are referred to here as simplex-truncated multivariate normal distributions (ST-MNDs).

For the original MND that was truncated on a unit simplex, and its equivalent lower-dimensional ST-MND, the values within the two mean vectors and covariance matrices of their corresponding non-truncated MNDs will generally differ from each other according to formulae provided in~\cite{Altmann2014}. However, \textit{after} the truncations are taken into account, the mean and covariance matrix elements associated with the non-redundant parameters in the original MND are \textit{equal} to the mean and covariance of the equivalent ST-MND. Thus, the mean and covariance of a ST-MND unambiguously defines the mean and covariance for non-redundant parameters of compositional data whose distributional shape otherwise follows a MND, so this mean vector and covariance matrix is likely of interest to compute. The mean and covariance of a ST-MND can be calculated from the mean and covariance of the corresponding non-truncated MND, although it is generally not easy to perform this calculation.

Similarly, probability densities on a MND truncated on a unit simplex are \textit{equal} to the probability densities on its equivalent lower-dimensional ST-MND. The probability density of a ST-MND within its truncated domain is greater than the probability density of the corresponding non-truncated MND, by a factor equal to the inverse of the integral of this non-truncated MND within the domain of the ST-MND. This integral, which is called here the integral of the ST-MND, depends only on the mean and covariance of the corresponding non-truncated MND, but is generally not easy to compute~\cite{Altmann2014,Koyama2020}. However, because knowledge of this integral's value permits calculation of probability densities for compositional data whose distributional shape otherwise follows a MND, the integral of the ST-MND is likely of interest to compute alongside its mean and covariance. In summary, rapid calculation of the integral, mean and covariance of a ST-MND from the mean and covariance of the corresponding non-truncated MND may be useful for a range of applications where compositional data is normally distributed within its allowable domain.

The easiest method for estimating the mean and covariance of a ST-MND is to calculate the sample mean and covariance of a large number of ST-MND samples, obtained by randomly sampling from the corresponding non-truncated MND and rejecting any samples that fall outside the domain of the ST-MND. This naive rejection sampling procedure can also be used to estimate the integral of the ST-MND  by computing the ratio of accepted samples to the total number of samples produced from the non-truncated MND. However, this method will become impractically slow if the acceptance ratio is very low (indicative of a very small integral).

Other methods that efficiently sample from the ST-MND even when the integral is small (reviewed in~\cite{Altmann2014}) can potentially provide faster estimates for the mean and covariance of this distribution. These methods are purposely designed to avoid computation of the integral. However, Gessner \textit{et al}.~\cite{Gessner2020} has recently pointed out that efficient sampling methods, in tandem with an appropriate method for estimating rare event probabilities \cite{Kroese2011}, can be used to efficiently estimate the integral of MNDs possessing linear domain constraints (of which ST-MNDs are a special case).

Another alternative and appealing approach for computing the integral, mean and covariance of truncated MNDs is to derive formulae for these three quantities in terms of integrals of hyperrectangularly-truncated MNDs, the latter of which can be readily computed in modern mathematical and statistical packages~\cite{Genz2009}. (Hyperrectangular truncations of MNDs should here be interpreted as the multivariate generalisation of rectangular truncations of bivariate normal distributions, with the upper and lower bounds of these hyperrectangles allowed to possess finite or infinite coordinates.) This approach has been used previously to compute the moments for MNDs truncated by constant lower and/or upper bounds in each dimension~\cite{Manjunath2021}, and for MNDs truncated by linear domain constraints for some covariance matrices~\cite{Lee2020}.

In this paper different approaches for estimating the integral, mean and covariance of any ST-MND are presented and compared. The three approaches compared are (1) naive rejection sampling, (2) the Gessner \textit{et al}.\ method described in~\cite{Gessner2020}, and (3) a semi-analytical method that uses new formulae derived for the integral, mean and covariance of a ST-MND in terms of integrals of hyperrectangularly-truncated MNDs. The main contribution of this work is therefore two-fold: firstly, to the author's best knowledge, it presents the first comparison of approaches for estimating the integral, mean and covariance of ST-MNDs, and secondly, one of the tested methods (the semi-analytical method) is a new and novel approach. All three tested methods are computationally compared for their speed and accuracy, and recommendations are provided for when each method may be the best. It is hoped that these recommendations can benefit any future modelling of compositional data whose fractions or probabilities would otherwise best be described via normal distributions.

\section*{Materials and methods}
The ``Materials and methods'' section is ordered as follows. The problem to be solved is mathematically defined, then separately each of the three approaches to solve the problem are described: naive rejection sampling, the Gessner \textit{et al}.~\cite{Gessner2020} method, and the semi-analytical method which is introduced in the present manuscript. Finally, details of the computational implementations used to compare the speed and accuracy of these three methods are provided.

\subsection*{Problem statement}
The multivariate normal distribution (MND) of dimension $n$, without any truncation, has probability density function 
\begin{equation}
\phi(\bm{x};\bm{\mu},\Sigma) = \dfrac{1}{\left(2 \pi\right)^{n/2} \left| \Sigma \right|^{1/2}} \exp \left( - \dfrac{1}{2} \left( \bm{x} - \bm{\mu} \right)^\top \Sigma^{-1} \left( \bm{x} - \bm{\mu} \right) \right),
\label{eq:phi}
\end{equation}
where the density function is evaluated at $\bm{x} \in \mathbb{R}^n$, and the distribution is characterised by mean vector $\bm{\mu} \in \mathbb{R}^n$ and a symmetric and positive semi-definite covariance matrix $\Sigma \in \mathbb{R}^{n \times n}$. It is denoted that $x_i$ and $\mu_i$ represent the $i$th elements of $\bm{x}$ and $\bm{\mu}$ respectively, and $\sigma_{ij}$ represents the $(i,j)$th element of $\Sigma$.\\

In the present work, the MND with truncations $x_i \geq 0$, $\forall i=1,..,n,$ and $\displaystyle\sum_{i=1}^n x_i \leq 1,$ is examined. This distribution, hereafter called the simplex-truncated multivariate normal distribution (ST-MND) and denoted as $\mathcal{N}_\T(\bm{\mu},\Sigma)$, has probability density function
\begin{equation}
\phi_\T(\bm{x};\bm{\mu},\Sigma) = \begin{cases} \dfrac{1}{Z} \phi(\bm{x};\bm{\mu},\Sigma), & \mbox{if } x_i \geq 0\mbox{, } \forall i=1,...,n, \mbox{ and } \displaystyle\sum_{i=1}^n x_i \leq 1, \\
0, & \mbox{otherwise,}\end{cases}
\label{eq:phi_T}
\end{equation}
where the integral $Z$ of this distribution is given by
\begin{equation}
Z = \int\displaylimits_{x_n=0}^1 \quad \int\displaylimits_{x_{n-1} = 0}^{1-x_n} \quad ... \quad \int\displaylimits_{x_2=0}^{1-\sum_{i=3}^n x_i} \quad \int\displaylimits_{x_1=0}^{1-\sum_{i=2}^n x_i} \phi(\bm{x};\bm{\mu},\Sigma) \,\, \dd x_1 \,\, \dd x_2 \,\, ... \,\, \dd x_{n-1} \,\, \dd x_n.
\label{eq:Z}
\end{equation}
Since $\int\phi(\bm{x};\bm{\mu},\Sigma) \, \dd \bm{x}$ is finite and $\phi(\bm{x};\bm{\mu},\Sigma)=|\phi(\bm{x};\bm{\mu},\Sigma)|$, Fubini's theorem is applicable to Eq~\eqref{eq:Z}. Thus, Eq~\eqref{eq:Z} with any $x_i$ and $x_j$ interchanged such that $i,j \in \left\{ 1,...,n \right\}$ is also a correct expression for $Z$. This paper characterises the distribution given by Eq~\eqref{eq:phi_T} by presenting methods to estimate its integral $Z \in (0,1)$, mean vector $\bm{\mu}_\T \in \mathbb{R}^n$ and covariance matrix $\Sigma_\T\in \mathbb{R}^{n \times n}$. Elements of $\bm{\mu}_\T \in \mathbb{R}^n$ and $\Sigma_\T\in \mathbb{R}^{n \times n}$ possess the following definitions. It is denoted that $(\mu_\T)_i$ represents the $i$th element of $\bm{\mu}_\T$, and $(\sigma_\T)_{ij}$ represents the $(i,j)$th element of $\Sigma_\T$, thus
\begin{equation} (\mu_\T)_i = \mathbb{E}(X_i), \qquad (\sigma_\T)_{ij} = \mathbb{E}(X_i X_j) - \mathbb{E}(X_i) \mathbb{E}(X_j), \qquad i,j=1,...,n,
\label{eq:elements}
\end{equation}
where $\mathbb{E}(\cdot)$ denotes an expectation, and $X_i$ and $X_j$ are the $i$th and $j$th elements respectively of the random vector $\bm{X} \sim \mathcal{N}_\T(\bm{\mu},\Sigma)$.\\

When the dimension of the ST-MND is one ($n=1$), Eq~\eqref{eq:phi_T} simplifies immediately to a univariate normal distribution with constant lower and upper bounds, so its integral, mean and covariance are already well-characterised elsewhere (e.g.~\cite{Johnson1994,Jawitz2004}). Therefore attention is restricted here to ST-MNDs with dimension $n \geq 2$. Analytical calculation of $Z$, $\bm{\mu}_\T$ and $\Sigma_\T$, whose elements are defined in Eq~\eqref{eq:Z} and Eq~\eqref{eq:elements}, is impossible for ST-MNDs with dimension $n \geq 2$. Hence, in the following sections three different methods that can estimate $Z$, $\bm{\mu}_\T$ and $\Sigma_\T$ are described.

\subsection*{Naive rejection sampling method}
The integral $Z$, mean $\bm{\mu}_\T$ and covariance matrix $\Sigma_\T$ of a ST-MND characterised by $\mathcal{N}_\T(\bm{\mu},\Sigma)$ can be estimated via naive rejection sampling as follows. The only user-specified requirement is the number of samples $M$ being used for the estimation of these quantities.

\begin{enumerate}
\item Set the number $m_{\mathrm{ST-MND}}$ of currently kept samples from the ST-MND to zero. Set the number $m_{\mathrm{total}}$ of currently trialled samples to zero.
\item Obtain a sample $\bm{x}$ from the corresponding non-truncated MND, $\bm{x} \sim \mathcal{N}(\bm{\mu},\Sigma)$. Set $m_{\mathrm{total}} \leftarrow m_{\mathrm{total}} + 1$.
\item If $\bm{x}$ obtained in Step 2 is in the domain of the ST-MND, i.e.\ if $x_i \geq 0$, $\forall i=1,..,n,$ and $\displaystyle\sum_{i=1}^n x_i \leq 1$, keep the sample and set $m_{\mathrm{ST-MND}} \leftarrow m_{\mathrm{ST-MND}} + 1$.
\item If the desired number of samples has been obtained, i.e.\ if $m_{\mathrm{ST-MND}} = M$, proceed to Step 5. Otherwise, go back to Step 2.
\item Estimate the mean $\bm{\mu}_\T$ and covariance matrix $\Sigma_\T$ from the sample mean and covariance of the $m_{\mathrm{ST-MND}}$ kept samples of $\bm{x}$.
\item Estimate the integral $Z$ via $Z \approx m_{\mathrm{ST-MND}}/m_{\mathrm{total}}$.
\end{enumerate}

Of the three methods described in this paper, naive rejection sampling is the easiest to implement. However, this algorithm suffers from large computational expense when the integral $Z$ is small, i.e.\ $Z \ll 1$. A small integral can occur if the non-truncated MND has a large covariance, or a mean that is far from the ST-MND domain, or if the distribution is of high dimension. This disadvantage motivates the usage of other algorithms for estimating the integral, mean and covariance of ST-MNDs.

\subsection*{Gessner \textit{et al}.\ method}
For the purpose of estimating the integral $Z$, mean $\bm{\mu}_\T$ and covariance matrix $\Sigma_\T$ of a ST-MND, the method proposed by Gessner \textit{et al}.~\cite{Gessner2020} can be used. This method itself consists of three algorithms: subset simulation~\cite{Au2001a}, the Holmes-Diaconis-Ross (HDR) algorithm~\cite{Kroese2011}, and an analytical version of elliptical slice sampling (ESS)~\cite{Murray2010}. The latter ESS method is called LIN-ESS~\cite{Gessner2020} to indicate its applicability to MNDs subject to \textit{linear} domain constraints (of which ST-MNDs are a special case). Subset simulation and the HDR algorithm are used to estimate the integral $Z$. LIN-ESS is used to perform rejection-free sampling in the ST-MND domain and thereby calculate a sample mean and covariance that approximates the mean $\bm{\mu}_\T$ and covariance $\Sigma_\T$ of the ST-MND.

Unification of all three of these algorithms is explained in~\cite{Gessner2020}, with their focus being on evaluating integrals of MNDs subject to linear domain constraints. Here these algorithms are summarised with specific emphasis on their implementation for finding the integral, mean and covariance of ST-MNDs. The description of LIN-ESS is provided first, because this method by itself can yield estimates of $\bm{\mu}_\T$ and $\Sigma_\T$, and because LIN-ESS is also used \textit{within} the subset simulation and HDR algorithms to estimate $Z$.

\subsubsection*{The LIN-ESS algorithm for a simplex-truncated multivariate normal distribution}
ESS is a Markov chain Monte Carlo algorithm to draw samples from a posterior when the prior distribution is a MND~\cite{Murray2010}. In the general form of ESS, the likelihood function depends on the problem at hand. In the more specific LIN-ESS algorithm, where the likelihood function needs only to represent linear domain constraints, the likelihood is assumed to output equal and non-zero values in all locations where all domain constraints are satisfied, and is zero at all other locations~\cite{Gessner2020}. In this section the application of LIN-ESS to sample from ST-MNDs is mathematically described.

ESS algorithms require that the MND has zero mean~\cite{Murray2010}, so the following coordinate transform from $\bm{x} \rightarrow \bm{y}$ is employed so that $\mathcal{N}_\T(\bm{\mu},\Sigma) \rightarrow \mathcal{N}_\T(\bm{0},\Sigma)$,
\begin{equation}
\bm{y} = \bm{x}-\bm{\mu}.
\label{eq:y}
\end{equation}
For LIN-ESS it is also convenient to rewrite the linear domain constraints in the form~\cite{Gessner2020}
\begin{equation}
A \bm{y} + \bm{c} \geq \bm{0}.
\label{eq:constraints}
\end{equation}
For a ST-MND of dimension $n$, there are $(n+1)$ constraints in total so Eq~\eqref{eq:constraints} is a column vector inequality defining $(n+1)$ separate inequalities. From comparison of Eq~\eqref{eq:y} and Eq~\eqref{eq:constraints} to the ST-MND domain constraints given in Eq~\eqref{eq:phi_T}, the following forms of the matrix $A$ and column vector $\bm{c}$ are readily derived for ST-MNDs. First, the matrix $A$ is of size $(n+1) \times n$, has its first row being a row vector of length $n$ containing $-1$ in all its elements (this row vector denoted below as $-\bm{1}_{1 \times n}$), and this row is concatenated vertically to the $n$-dimensional identity matrix (denoted below as $I_n$),
\begin{equation}
A = \begin{bmatrix} -\bm{1}_{1 \times n} \\ I_n \end{bmatrix}.
\label{eq:constraint_A}
\end{equation}
Second, the column vector $\bm{c}$ is of length $(n+1)$, with its uppermost element equal to unity minus the sum of all non-truncated MND means, concatenated vertically to the mean vector $\bm{\mu}$ of the non-truncated MND,
\begin{equation}
\bm{c} = \begin{bmatrix} 1 - \displaystyle\sum_{i=1}^n \mu_i \\ \bm{\mu} \end{bmatrix}.
\label{eq:constraint_c}
\end{equation}\\

The LIN-ESS algorithm for sampling from ST-MNDs then proceeds as follows. Starting from a randomly chosen location $\bm{y}$ within the ST-MND domain, denoted as $\bm{y}_0$, each successive sample $\bm{y}_1$, $\bm{y}_2$, etc.\ from the ST-MND is a location within its domain on a sampled ellipse that the previous sample also falls on. Hence, the LIN-ESS algorithm for ST-MNDs only requires definition of the rule for moving from the previous sample $\bm{y}_t$ along an appropriately sampled ellipse to find the next sample $\bm{y}_{t+1}$. Specifically, the LIN-ESS algorithm's procedure for choosing the next sample location is as follows:
\begin{enumerate}
\item Sample an auxiliary vector $\bm{\nu} \sim \mathcal{N}(\bm{0},\Sigma)$, where $\Sigma$ is the covariance matrix of the MND.
\item Construct the ellipse $\bm{y}^*(\theta)$ satisfying
\begin{equation}
\bm{y}^*(\theta) = \bm{y}_t \cos \theta + \bm{\nu} \sin \theta,\quad \mbox{where} \quad 0 \leq \theta < 2 \pi.
\label{eq:ellipse}
\end{equation}
\item Identify all values of $\theta$ on the ellipse that fall within the ST-MND domain.
\item Define a uniform probability distribution over the $\theta$ values identified in the previous step, randomly sample a value of $\theta$ from this distribution, and denote this sampled value as $\theta^*$.
\item Place the next sample $\bm{y}_{t+1}$ at $\bm{y}^*(\theta^*)$.
\end{enumerate}

\begin{figure}[!h]
\includegraphics[width=\textwidth]{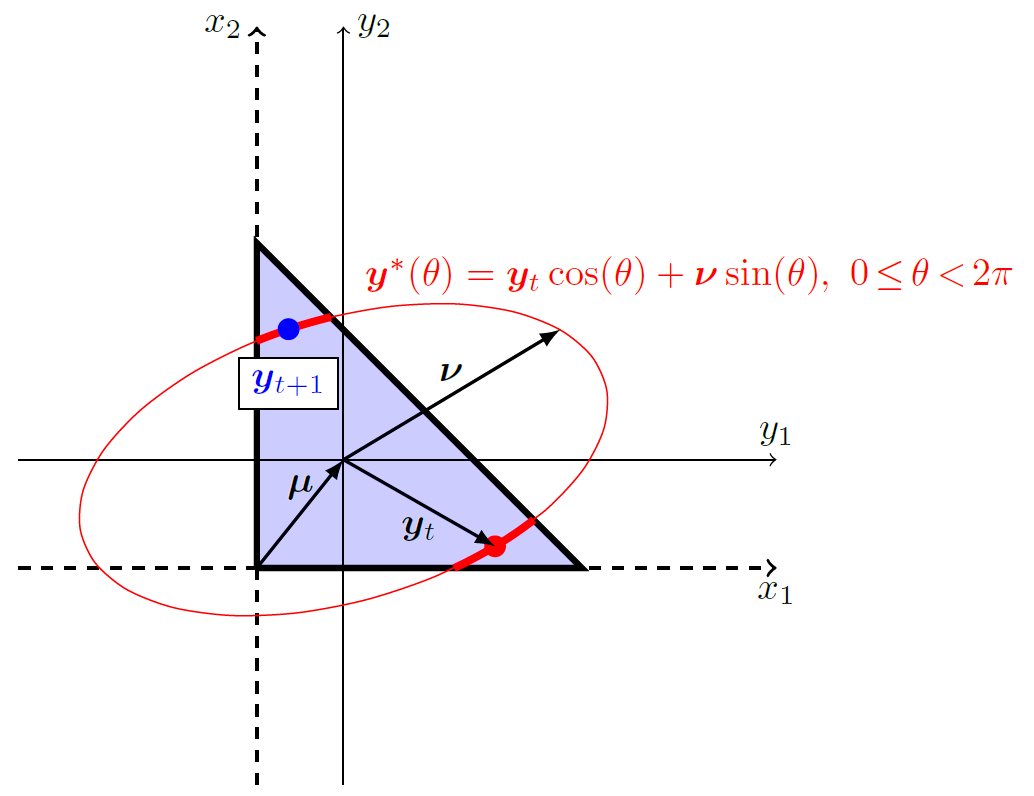}
\caption{{\bf Graphical illustration of sampling from a simplex-truncated bivariate normal distribution $\mathcal{N}_{\T}(\bm{\mu},\bm{\Sigma})$ using the LIN-ESS algorithm.} Coordinates are transformed from $(x_1,x_2)$ (thin black dashed lines) to $(y_1,y_2)$ (thin black solid lines) so that the origin is located at the mean $\bm{\mu}$ of the distribution. With the previous sample's location indicated by $\bm{y}_t$, an ellipse $\bm{y}^*(\theta),\,0$$\,\leq\,$$\theta$$\,<\,$$2\pi$ (shown in red) is obtained from $\bm{y}_t$ and sampling a vector $\bm{\nu} \sim \mathcal{N}(0,\bm{\Sigma})$. Values of $\theta$ on arcs of the ellipse that fall within the distribution's domain (thick red arcs) are identified, and one value $\theta^*$ is randomly sampled from a uniform distribution on the range of $\theta$ values that fall within the domain. The resulting location $\bm{y}^*(\theta^*)$ becomes the next sample $\bm{y}_{t+1}$ from the distribution. Notice that this sampling procedure has a 100\% acceptance rate. Figure adapted from~\cite{Gessner2020}.}
\label{fig:ESS_Simplex}
\end{figure}

These algorithmic steps are illustrated in Fig~\ref{fig:ESS_Simplex}, for sampling from a ST-MND of dimension $n=2$. In two dimensions, the ST-MND has the constraints $x_1 \geq 0$, $x_2 \geq 0$ and $x_1+x_2 \leq 1$, so the domain is bounded by a right-angled triangle with coordinates $(0,0)$, $(0,1)$ and $(1,0)$ in the original ($\bm{x}$) coordinate system. However, for ESS algorithms the MND must have zero mean, so the origin of the transformed ($\bm{y}$) coordinate system is instead located at the mean $\bm{\mu}$ of the MND. From the sample's previous location $\bm{y}_t$ (red dot in Fig~\ref{fig:ESS_Simplex}) and the sampled auxiliary vector $\bm{\nu}$ (Step 1 listed above), an ellipse $\bm{y}^*(\theta)$ is drawn (Step 2; thin red ellipse in Fig~\ref{fig:ESS_Simplex}). All values of $\theta$ on the ellipse that fall within the domain are identified (Step 3), which in the illustrated case comprises $\theta$ values on two arcs of the ellipse (thick red arcs in Fig~\ref{fig:ESS_Simplex}). A value of $\theta$ from these arcs is then chosen at random (Steps 4 and 5) to be the next sample $\bm{y}_{t+1}$ from the ST-MND (blue dot in Fig~\ref{fig:ESS_Simplex}).

A clear advantage of the LIN-ESS algorithm over the naive rejection sampling algorithm is that there is no rejection of samples. When moving from one sample $\bm{y}_t$ along an ellipse to the next sample $\bm{y}_{t+1}$, the LIN-ESS algorithm is guaranteed to identify a new suitable sample location $\bm{y}_{t+1}$ so long as values of $\theta$ on this ellipse within the allowable domain can be identified. Supplementary Material S1 provides the mathematical details required to identify all values of $\theta$ on the ellipse that fall within the ST-MND domain (which is Step 3 of the LIN-ESS algorithm's procedure for choosing the next sample location). The mathematical procedure provided in Supplementary Material S1 is applicable for identifying values of $\theta$ that fall within \textit{any} domain with linear constraints written as the vector inequality given in Eq~\eqref{eq:constraints}. In other words, for this mathematical procedure the constraints do not necessarily need to possess the values of $A$ and $\bm{c}$ that are specified for ST-MNDs in Eq~\eqref{eq:constraint_A} and Eq~\eqref{eq:constraint_c}. The procedure provided in Supplementary Material S1 is therefore very similar to that described in~\cite{Gessner2020}. However, some different trigonometric relations are used in the mathematical procedures here, so these relations are described carefully in Supplementary Material S1.\\

Finally, it is worth remarking that LIN-ESS is a Markov chain Monte Carlo algorithm, and thus consecutive samples from this algorithm are correlated. To address this issue, the present implementation primarily followed~\cite{Gessner2020} by keeping every second sample from LIN-ESS for estimation of the ST-MND mean and covariance. Once a sufficient number of samples are obtained, their sample mean and covariance can be used as an estimate of the mean $\bm{\mu}_\T$ and covariance $\Sigma_\T$ of the ST-MND.

\subsubsection*{Subset simulation and the Holmes-Diaconis-Ross algorithm}
The integral $Z$ of a ST-MND can be estimated using the combination of subset simulation~\cite{Au2001a}, the HDR algorithm~\cite{Kroese2011} and an appropriate method for sampling MNDs from domains truncated by linear constraints such as LIN-ESS~\cite{Gessner2020}. Implementations of subset simulation and the HDR algorithm for this purpose are very similar, so are described here together. Subset simulation by itself can yield a biased estimate of $Z$. However, an unbiased estimate of $Z$ may be accurately and efficiently obtained by using the outcomes of subset simulation to inform the HDR algorithm.

First, the unrestricted $n$-dimensional domain $\mathbb{R}^n$ of the non-truncated MND is denoted as $\mathcal{L}_0$, and the truncated domain of the ST-MND is denoted as $\mathcal{L}$. A set of $s=1,...,S$ nested domains $\mathcal{L}_s$ is sought, such that
\begin{equation}
\mathcal{L} = \mathcal{L}_S \subset \mathcal{L}_{S-1} \subset ... \subset \mathcal{L}_2 \subset \mathcal{L}_1 \subset \mathcal{L}_0 = \mathbb{R}^n.
\end{equation}
Thus, if $p(\mathcal{L}_s | \mathcal{L}_{s-1})$ denotes the conditional probability of a location being present in the domain $\mathcal{L}_s$ given that this location is also in the next largest domain $\mathcal{L}_{s-1}$, the integral $Z$ of the ST-MND can be calculated via
\begin{equation}
Z = p(\mathcal{L}_0) \displaystyle\prod_{s=1}^S p(\mathcal{L}_s | \mathcal{L}_{s-1}).
\label{eq:nested}
\end{equation}
For ST-MNDs, $p(\mathcal{L}_0)=1$ since the integral of a non-truncated MND is unity. For numerical implementations of calculating $Z$ using Eq~\eqref{eq:nested}, this equation is typically written in logarithmic form so that $\log(Z)$ can be calculated from a sum of logarithms of probabilities.

If it is possible to sample from each nested domain $\mathcal{L}_s$, then each conditional probability $p(\mathcal{L}_s | \mathcal{L}_{s-1})$ in Eq~\eqref{eq:nested} can be approximated from the proportion of samples from domain $\mathcal{L}_{s-1}$ that also fall within the domain $\mathcal{L}_s$. Since LIN-ESS (see previous section) can yield samples from a MND truncated by any set of linear domain constraints~\cite{Gessner2020}, LIN-ESS can therefore be used to approximate all values of $p(\mathcal{L}_s | \mathcal{L}_{s-1})$ if each nested domain $\mathcal{L}_s$ is truncated only by linear domain constraints. A convenient choice for such nested domains $\mathcal{L}_s$ is to require them to satisfy loosened ST-MND domain constraints, with this loosening defined by a non-negative shift value $\gamma_s$,
\begin{equation}
A \bm{y} + \bm{c}' \geq \bm{0}, \quad \mbox{where} \quad A = \begin{bmatrix} -\bm{1}_{1 \times n} \\ I_n \end{bmatrix}, \quad
\bm{c}' = \bm{c} + \gamma_s \bm{1}_{(n+1) \times 1}, \quad \bm{c} = \begin{bmatrix} 1 - \displaystyle\sum_{i=1}^n \mu_i \\ \bm{\mu} \end{bmatrix}.
\label{eq:constraints_gamma}
\end{equation}
These constraints differ from the ST-MND constraints \eqref{eq:constraints}-\eqref{eq:constraint_c} because the vector $\bm{c}$ is replaced by $\bm{c}'$, and $\bm{c}'$ differs from $\bm{c}$ only by the addition of $\gamma_s$ within all of its elements. Using this form of the linear constraints for nested domains $\mathcal{L}_s$, the non-negative shift values $\gamma_s$ decrease as the domain becomes more restrictive (i.e.\ as $s$ increases). Also, the shift value becomes zero for the domain $\mathcal{L}_S = \mathcal{L}$ so that constraints in Eq~\eqref{eq:constraints_gamma} applied to this most restrictive domain match those of the ST-MND indicated in Eq~\eqref{eq:constraints}-\eqref{eq:constraint_c}.

If sampling from each of the nested domains is used to estimate the conditional probabilities $p(\mathcal{L}_s | \mathcal{L}_{s-1})$, the subsequent computation of $Z$ using Eq~\eqref{eq:nested} may be both accurate and efficient if all values of $p(\mathcal{L}_s | \mathcal{L}_{s-1})$ are far from both zero and one. Subset simulation uses a target value $\rho$ for $p(\mathcal{L}_s | \mathcal{L}_{s-1})$ to adaptively choose all values of $\gamma_s$ that define the nested domains $\mathcal{L}_s$. Potential choices for $\rho$ include 0.1~\cite{Au2001b} and 0.5~\cite{Gessner2020}. However, this use of a target value $\rho$ to adaptively construct each nested domain $\mathcal{L}_s$ also causes Eq~\eqref{eq:nested} to yield biased estimates of $Z$.\\

The HDR algorithm proceeds similarly to subset simulation except that the nested domains are pre-selected, so that the estimates of $Z$ obtained from Eq~\eqref{eq:nested} are unbiased. However, if the nested domains are chosen poorly, conditional probabilities $p(\mathcal{L}_s | \mathcal{L}_{s-1})$ may be close to zero or one, causing inaccuracies and/or inefficiencies for estimating $p(\mathcal{L}_s | \mathcal{L}_{s-1})$ using sample-based methods. This issue can be resolved by using subset simulation to identify suitable values of $\gamma_s$ to characterise the nested domains, and then to use these nested domains in the HDR algorithm to obtain an unbiased estimate of $Z$, thus maximising the strengths of both methods~\cite{Gessner2020}.

Combining all of the above considerations, unbiased estimation of the integral $Z$ of a ST-MND using subset simulation and the HDR algorithm can proceed as follows.

\paragraph{Subset simulation:}
\begin{enumerate}
\item Obtain $M$ samples of the non-truncated MND, via $\bm{y}_m \sim \mathcal{N}(\bm{0},\Sigma)$, $\forall m=1,...,M$. (Note that the MND being sampled here has a mean of zero because the sampling occurs in the transformed coordinates $\bm{y}$ that are centered on the mean $\bm{\mu}$ of this distribution, see Fig~\ref{fig:ESS_Simplex}.)
\item Initialise $\log(Z) = 0$, and set the current nested domain number as $s=0$.
\item Calculate the individual shift values $g_m$ of all $M$ current samples, as $g_m = -\min \left\{ A \bm{y}_m + \bm{c} \right\}$, $\forall m=1,...,M$, where $A$ and $\bm{c}$ are defined by Eq~\eqref{eq:constraint_A} and Eq~\eqref{eq:constraint_c}.
\item Choose the shift value $\gamma_{s+1}$ of the $(s+1)$th nested domain so that a proportion $\rho$ of the $M$ current samples possess $g_m < \gamma_{s+1}$. If the obtained number for $\gamma_{s+1}$ is negative, set $\gamma_{s+1}=0$.
\item Update $\log(Z) \leftarrow \log(Z) + \log \left(\left.M\right|_{g_m < \gamma_{s+1}} \right) - \log(M)$, where $\left.M\right|_{g_m < \gamma_{s+1}}$ is the number of current samples that possess an individual shift value satisfying $g_m < \gamma_{s+1}$.
\item If $\gamma_{s+1}=0$, conclude the algorithm. Otherwise, go to Step 7.
\item Set the current nested domain number $s \leftarrow s+1$, and discard all previous samples.
\item Obtain $M$ samples of the MND $\mathcal{N}(\bm{0},\Sigma)$ truncated by the linear constraints shown in Eq~\eqref{eq:constraints_gamma} using any suitable sampling method (e.g.\ LIN-ESS), and go back to Step 3.
\end{enumerate}
\paragraph{Holmes-Diaconis-Ross algorithm:} This algorithm is almost an exact repeat of subset simulation. The only required change is to replace Step 4 by the following:
\begin{enumerate}
\item[4.] Obtain $\gamma_{s+1}$ from its value obtained in subset simulation.
\end{enumerate}

The present implementations of these algorithms used the settings recommended by Gessner \textit{et al}.~\cite{Gessner2020}, as follows. To maximise the entropy of the binary distribution over whether samples fall inside or outside the next nested domain, the conditional probability target value was set to $\rho = 0.5$. The number of samples in subset simulation was set to $M=16$, to rapidly obtain a reasonable sequence of nested domains, but because of the dependency of the subset construction on the samples themselves, only every tenth sample was kept from LIN-ESS to obtain these samples~\cite{Gessner2020}.  Finally, for the HDR algorithm, every second sample was kept and a much larger number of samples $M$ was used for accurate computation of the ST-MND integral.

\subsection*{Semi-analytical method}
\label{sec:SAM}

An alternative approach for estimating the integral $Z$, mean $\bm{\mu}_\T$ and covariance $\Sigma_\T$ of a ST-MND, that does not require user implementation of sampling methods, is to mathematically express these three quantities in terms of integrals of hyperrectangularly-truncated MNDs. In other words, this approach seeks to express $Z$, $\bm{\mu}_\T$ and $\Sigma_\T$ of a ST-MND in terms of integrals $\Phi(\bm{a},\bm{b};\bm{\mu},\Sigma)$ of the form
\begin{equation}
\Phi(\bm{a},\bm{b};\bm{\mu},\Sigma) = \displaystyle\int_{\bm{a}}^{\bm{b}} \phi(\bm{x};\bm{\mu},\Sigma) \,\, \dd \bm{x}, \qquad \bm{x}, \bm{\mu}, \bm{a}, \bm{b} \in \mathbb{R}^n, \qquad \Sigma \in \mathbb{R}^{n \times n}.
\label{eq:hyper}
\end{equation}
In Eq~\eqref{eq:hyper}, $\bm{a}$ and $\bm{b}$ define the hyperrectangular truncation $\bm{a} \leq \bm{x} \leq \bm{b}$ of the MND. Comparison of Eq~\eqref{eq:Z} and Eq~\eqref{eq:hyper} reveals that the integral of a ST-MND of dimension $n \geq 2$ contains dependence of its integration limits on $\bm{x}$, and thus this integral $Z$ does not adhere to the form of Eq~\eqref{eq:hyper}. The approach introduced in this paper of using mathematical relationships to describe $Z$, $\bm{\mu}_\T$ and $\Sigma_\T$ in terms of integrals of the form of Eq~\eqref{eq:hyper} is referred to here as the semi-analytical method.

Integrals given in Eq~\eqref{eq:hyper} typically do not have closed form, but can be readily computed in modern mathematical and statistical packages~\cite{Genz2009}. For example, these integrals can be estimated by use of the function ``mvncdf'' in MATLAB, and the function ``pmvnorm'' in R. A potential advantage of the semi-analytical method is that its computations may be very rapid, if evaluation of the integrals given in Eq~\eqref{eq:hyper} is already well-optimised within the software package being used.

In the following sections, the semi-analytical method for expressing $Z$, $\bm{\mu}_\T$ and $\Sigma_\T$ in terms of integrals of the form given in Eq~\eqref{eq:hyper} is derived. This approach combines usage of the inclusion-exclusion principle for convex polyhedra~\cite{Edelsbrunner1995}, a coordinate transformation suitable for some MNDs with linear constraints~\cite{Geweke1991}, and moment calculations for hyperrectangularly-truncated MNDs~\cite{Manjunath2021}. First, the application of the semi-analytical method to simplex-truncated bivariate normal distributions (i.e.\ ST-MNDs with dimension $n=2$) is described separately for its integral, mean and covariance. The extension of the semi-analytical method to the general case (ST-MNDs of any dimension $n$) is then summarised.

\subsubsection*{Semi-analytical method: Integral of the simplex-truncated bivariate normal distribution}
\label{sec:SAM_Bivariate_Integral}

To calculate the integral of a ST-MND using the semi-analytical method, two steps are required. First, this integral is expressed in terms of integrals of MNDs that are truncated only by a number of linear domain constraints equal to or less than their dimension. Second, using a suitable coordinate transformation, the latter truncated MNDs are written in the form of hyperrectangularly-truncated MNDs so that their integrals can be readily evaluated in modern software.

To accomplish the first step, note that every $n$-dimensional ST-MND is truncated by $(n+1)$ constraints. In other words, the number of constraints in a ST-MND is always one more than its dimension. For example, the simplex-truncated bivariate normal distribution is a distribution within $\mathbb{R}^2$ that is truncated by three constraints: $x_1 \geq 0$, $x_2 \geq 0$ and $x_1 + x_2 \leq 1$. However, the integral over this triangular region can be rewritten in terms of integrals over MNDs that are each truncated by zero, one or two constraints~\cite{Koyama2020}. To accomplish this, three spatial domains $\mathcal{L}_1$, $\mathcal{L}_2$ and $\mathcal{L}_3$ are defined, all of which are half-spaces of $\mathbb{R}^2$ that are the regions of space ``sliced off'' to yield the desired truncations $x_1 \geq 0$, $x_2 \geq 0$ and $x_1 + x_2 \leq 1$ respectively:
\begin{equation}
\begin{array}{l}
\mbox{Domain } \mathcal{L}_1: \mbox{All } \bm{x} \mbox{ satisfying } x_1 < 0 \\
\mbox{Domain } \mathcal{L}_2: \mbox{All } \bm{x} \mbox{ satisfying } x_2 < 0 \\
\mbox{Domain } \mathcal{L}_3: \mbox{All } \bm{x} \mbox{ satisfying } x_1 + x_2 > 1
\label{eq:regions}
\end{array}
\end{equation}

\begin{figure}[!h]
\includegraphics[width=\textwidth]{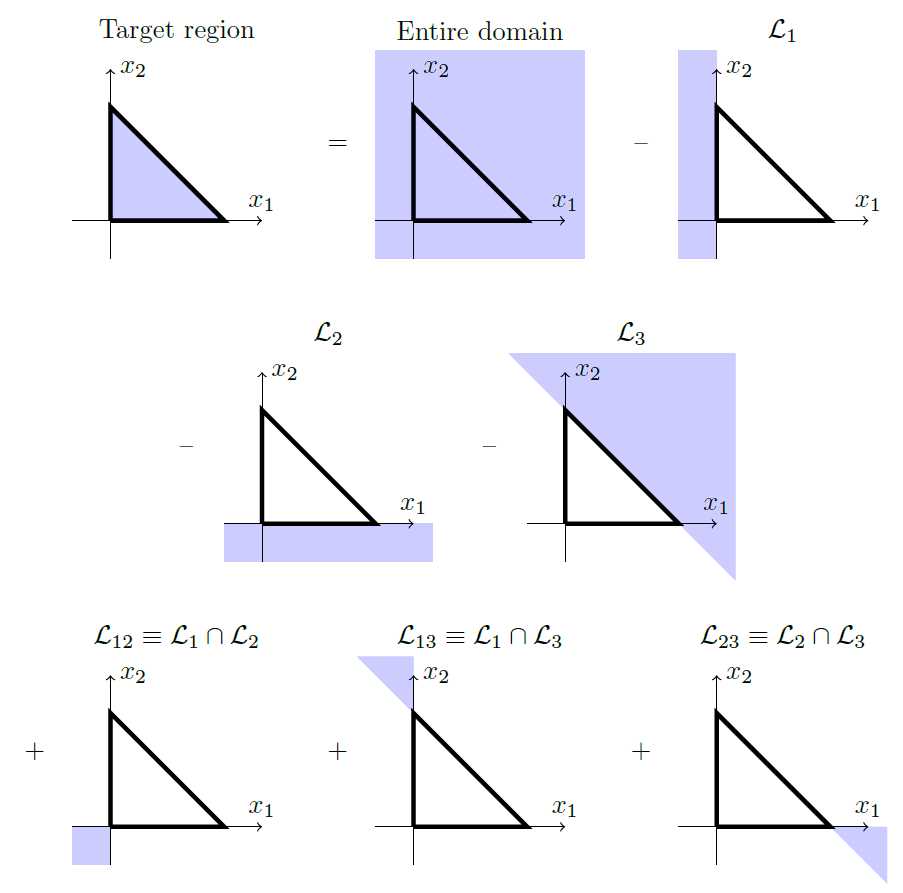}
\caption{{\bf Application of the inclusion-exclusion principle to express the target region enclosed by the non-negative space under a simplex in two dimensions ($x_1 \geq 0$, $x_2 \geq 0$, $x_1+x_2\leq 1$) in terms of domains that are truncated by no more than two constraints.} Figure adapted from \cite{Koyama2020}.}
\label{fig:triangle}
\end{figure}

It is convenient here to introduce the notation that multiple indices $\bm{v} = ijk...$ of a domain $\mathcal{L}_{\bm{v}}$ indicate that the domain is formed by the intersection of multiple half-space domains, e.g.\ $\mathcal{L}_{ij} \equiv \mathcal{L}_i \cap \mathcal{L}_j$. As shown in Fig~\ref{fig:triangle}, the definitions of the half-space domains $\mathcal{L}_1$, $\mathcal{L}_2$ and $\mathcal{L}_3$ given in Eq~\eqref{eq:regions} permit the content of the two-dimensional non-negative space bounded under a unit simplex to be defined in terms of the content of spatial domains which are each truncated by no more than two constraints. This is an example application of the inclusion-exclusion principle~\cite{Edelsbrunner1995}, so-called because the estimation of the target region requires over-generous inclusion, followed by a sequence of compensating exclusions and inclusions. For example, in Fig~\ref{fig:triangle}, the initial inclusion (the entire domain) is an overestimate of the target region; this inclusion is compensated by excluding the domains $\mathcal{L}_1$, $\mathcal{L}_2$ and $\mathcal{L}_3$. However, these exclusions are again too generous so require subsequent inclusion of the domains $\mathcal{L}_{12}$, $\mathcal{L}_{13}$ and $\mathcal{L}_{23}$.

For the simplex-truncated bivariate normal distribution characterised by probability density $\phi_\T(\bm{x};\bm{\mu},\Sigma)$, the inclusion-exclusion principle therefore indicates that this distribution's integral $Z$ can be expressed as
\begin{equation}
Z = 1 - \Phi_1 - \Phi_2 - \Phi_3 + \Phi_{12} + \Phi_{13} + \Phi_{23},
\label{eq:Z_expression}
\end{equation}
where $\Phi_{\bm{v}}$ represents the integral of $\phi(\bm{x};\bm{\mu},\Sigma)$ over the domain $\mathcal{L}_{\bm{v}}$. Eq~\eqref{eq:Z_expression} gives that, for the bivariate case, six integrals of MNDs truncated by linear domain constraints need to be evaluated. All of these integrals possess a number of constraints (one or two) that is equal to or less than the dimension (two) of the spatial region.

The second step in calculating $Z$ using the semi-analytical method is to transform the MNDs that are truncated only by a number of linear domain constraints equal to or less than their dimension, to hyperrectangularly-truncated MNDs. As pointed out in \cite{Geweke1991}, if the probability density $\phi_L(\bm{x};\bm{\mu},\Sigma)$ of a MND of dimension $n$ truncated by no more than $n$ linear domain constraints is written in the form
\begin{equation}
\phi_L(\bm{x};\bm{\mu},\Sigma) \propto \begin{cases} \phi(\bm{x};\bm{\mu},\Sigma), & \mbox{if } \bm{c} < T \bm{x} < \bm{d}, \\
0, & \mbox{otherwise,}\end{cases}
\label{eq:phi_L}
\end{equation}
where the constraint vectors $\bm{c}, \bm{d} \in \mathbb{R}^n$, and the transformation matrix $T \in \mathbb{R}^{n \times n}$ has been defined such that its rank is equal to its dimension, this distribution can be transformed to an equivalent MND with
zero mean and hyperrectangular truncation, possessing probability density $\phi_R(\bm{w};\bm{0},\varepsilon)$ given by
\begin{equation}
\phi_R(\bm{w};\bm{0},\varepsilon) \propto \begin{cases} \phi(\bm{w};\bm{0},\varepsilon), & \mbox{if } \bm{a} < \bm{w} < \bm{b}, \\
0, & \mbox{otherwise,}\end{cases}
\label{eq:phi_R}
\end{equation}
where
\begin{equation}
\bm{x} = T^{-1} \bm{w} + \bm{\mu}, \qquad \varepsilon = T \Sigma T^\top, \qquad \bm{a} = \bm{c} - T \bm{\mu}, \qquad 
\bm{b} = \bm{d}- T \bm{\mu}.
\label{eq:transform}
\end{equation}
Since the integral of Eq~\eqref{eq:phi_R} is of the form given in Eq~\eqref{eq:hyper}, the transformation described in Eq~\eqref{eq:transform} allows each of the six integrals $\Phi_{\bm{v}}$ in Eq~\eqref{eq:Z_expression} to be readily computed as
\begin{equation}
\Phi_{\bm{v}} = \Phi(\bm{c}_{\bm{v}} - T_{\bm{v}} \bm{\mu}, \,\, \bm{d}_{\bm{v}}-T_{\bm{v}} \bm{\mu}; \,\, \bm{0}, \,\, T_{\bm{v}} \Sigma T_{\bm{v}}^\top).
\label{eq:phi_v}
\end{equation}
All that is needed are expressions for $T_{\bm{v}}$, $\bm{c}_{\bm{v}}$ and $\bm{d}_{\bm{v}}$, so that the rank of all $T_{\bm{v}}$ is two, for each of the six integrals in Eq~\eqref{eq:Z_expression}. Expressions for these matrices and vectors, which can be readily deduced from the inequalities on $\bm{x}$ that bound the required integrals, are provided in Table~\ref{table:R}.

\begin{table}[!ht]
\centering
\caption{{\bf Transformation matrices $T_{\bm{v}}$, and constraint vectors $\bm{c}_{\bm{v}}$ and $\bm{d}_{\bm{v}}$, that define the truncation $\bm{c}_{\bm{v}} < T_{\bm{v}} \bm{x} < \bm{d}_{\bm{v}}$ for six relevant regions shown in Fig~\ref{fig:triangle}.} Dummy inequalities $-\infty < x_i < \infty$ were introduced into the transformation matrices $T_1$, $T_2$ and $T_3$ to ensure that these matrices have rank equal to their dimension. The order of rows in $T_{23}$, $\bm{c}_{23}$ and $\bm{d}_{23}$ has been switched so that this matrix and these vectors possess the form that would be obtained from a generalised two-step algorithm for ST-MNDs of any dimension described later in the ``Materials and methods'' section.}
\label{table:R}
\begin{tabular}{cccccc}
\\ \hline
Region & $\bm{v}$ & Constraints & $T_{\bm{v}}$ & $\bm{c}_{\bm{v}}$ & $\bm{d}_{\bm{v}}$ \\
\hline
$\mathcal{L}_1$ & 1 & $x_1 < 0$ & $\begin{bmatrix} 1 & 0 \\ 0 & 1 \end{bmatrix}$ & $\begin{bmatrix} -\infty \\ -\infty \end{bmatrix}$ & $\begin{bmatrix} 0 \\ \infty \end{bmatrix}$ \\ \\
$\mathcal{L}_2$ & 2 & $x_2 < 0$ & $\begin{bmatrix} 1 & 0 \\ 0 & 1 \end{bmatrix}$ & $\begin{bmatrix} -\infty \\ -\infty \end{bmatrix}$ & $\begin{bmatrix} \infty \\ 0 \end{bmatrix}$ \\ \\
$\mathcal{L}_3$ & 3 & $x_1 + x_2 > 1$ & $\begin{bmatrix} 1 & 1 \\ 1 & 0 \end{bmatrix}$ & $\begin{bmatrix} 1 \\ -\infty \end{bmatrix}$ & $\begin{bmatrix} \infty \\ \infty \end{bmatrix}$ \\ \\
$\mathcal{L}_{12} \equiv \mathcal{L}_1 \cap \mathcal{L}_2$ & 12 & $x_1 < 0$, \, $x_2 < 0$ & $\begin{bmatrix} 1 & 0 \\ 0 & 1 \end{bmatrix}$ & $\begin{bmatrix} -\infty \\ -\infty \end{bmatrix}$ & $\begin{bmatrix} 0 \\ 0 \end{bmatrix}$ \\ \\
$\mathcal{L}_{13} \equiv \mathcal{L}_1 \cap \mathcal{L}_3$ & 13 & $x_1 < 0$, \, $x_1+x_2 > 1$ & $\begin{bmatrix} 1 & 0 \\ 1 & 1 \end{bmatrix}$ & $\begin{bmatrix} -\infty \\ 1 \end{bmatrix}$ & $\begin{bmatrix} 0 \\ \infty \end{bmatrix}$ \\ \\
$\mathcal{L}_{23} \equiv \mathcal{L}_2 \cap \mathcal{L}_3$ & 23 & $x_1+x_2 > 1$, \, $x_2 < 0$  & $\begin{bmatrix} 1 & 1 \\ 0 & 1 \end{bmatrix}$ & $\begin{bmatrix} 1 \\ -\infty \end{bmatrix}$ & $\begin{bmatrix} \infty \\ 0 \end{bmatrix}$ \\ \\
\hline
\end{tabular}
\end{table}

\subsubsection*{Semi-analytical method: Mean of the simplex-truncated bivariate normal distribution}
To calculate the mean of a ST-MND using the semi-analytical method, three steps are required. First, this mean is expressed in terms of means and integrals of MNDs that are truncated only by a number of linear domain constraints equal to or less than their dimension. Second, the means and integrals of the latter truncated MNDs are expressed in terms of means and integrals of hyperrectangularly-truncated MNDs via the same coordinate transformation~\cite{Geweke1991} that was used in the previous section. Third, the means of these hyperrectangularly-truncated MNDs are expressed in terms of integrals of hyperrectangularly-truncated MNDs using moment formulae derived in~\cite{Manjunath2021}.\\

To accomplish the first step for calculating the mean of a ST-MND, the inclusion-exclusion principle can be used together with the observation that expectations (such as the desired mean vector $\bm{\mu}_\T$) can be written as sums of conditional expectations multiplied by the probabilities of these conditions. For the bivariate case, the inclusion-exclusion principle shown in Fig~\ref{fig:triangle} thus implies that elements $(\mu_{\T})_i$, $i=1,2,$ of the mean vector $\bm{\mu}_\T$, satisfy
\begin{multline}
(\mu_{\T})_i = \dfrac{1}{Z} \left( \mu_i - \Phi_1 \mu_{i,1} - \Phi_2 \mu_{i,2} - \Phi_3 \mu_{i,3} \right. \\ \left. + \, \Phi_{12} \mu_{i,12} + \Phi_{13} \mu_{i,13} + \Phi_{23} \mu_{i,23} \right), \qquad i=1,2, \qquad
\label{eq:mu_T_conditional}
\end{multline}
where $\mu_{i,\bm{v}}$ is the expectation $\mathbb{E}(X_i)$ for the bivariate normal distribution $\mathcal{N}(\bm{\mu},\Sigma)$ truncated within the domain $\mathcal{L}_{\bm{v}}$ (see Table~\ref{table:R}). 

For the second step of calculating the mean of a ST-MND, $\mu_i$ is known, all $\Phi_{\bm{v}}$ and $Z$ are calculated as in the previous section, and expressions for $\mu_{i,\bm{v}}$ are obtained as follows. Since  all $\mu_{i,\bm{v}}$ are means for truncated MNDs of the form given by Eq~\eqref{eq:phi_L}, they can all be transformed to means for truncated MNDs of the form given by Eq~\eqref{eq:phi_R} using the transformation described by Eq~\eqref{eq:transform}. Using the notation $\mathbb{E}(f(\bm{W}); \bm{0}, \varepsilon, \bm{a}, \bm{b})$ to denote the expectation of some function $f(\bm{W})$ for a truncated MND whose probability density is given by Eq~\eqref{eq:phi_R}, each $\mu_{i,\bm{v}}$ can therefore be expressed as
\begin{equation}
\mu_{i,\bm{v}} = \mathbb{E}\left( X_i ; \,\, \bm{0}, \,\, T_{\bm{v}} \Sigma T_{\bm{v}}^\top, \,\, \bm{c}_{\bm{v}} - T_{\bm{v}} \bm{\mu}, \,\, \bm{d}_{\bm{v}} - T_{\bm{v}} \bm{\mu} \right),
\label{eq:mu_iv_before}
\end{equation}
where the transformation matrices $T_{\bm{v}}$ and constraint vectors $\bm{c}_{\bm{v}}$, $\bm{d}_{\bm{v}}$ for the bivariate case are provided in Table~\ref{table:R}. Noting from Eq~\eqref{eq:transform} that $\bm{x} = T_{\bm{v}}^{-1} \bm{w} + \bm{\mu}$, Eq~\eqref{eq:mu_iv_before} can be rewritten for the bivariate case as
\begin{equation}
\mu_{i,\bm{v}} = \mu_i + \displaystyle\sum_{j=1}^2 (T_{\bm{v}})^{-1}_{ij} \mathbb{E}\left( W_j ; \,\, \bm{0}, \,\, T_{\bm{v}} \Sigma T_{\bm{v}}^\top, \,\, \bm{c}_{\bm{v}} - T_{\bm{v}} \bm{\mu}, \,\, \bm{d}_{\bm{v}} - T_{\bm{v}} \bm{\mu} \right),
\label{eq:mu_iv}
\end{equation}
where $(T_{\bm{v}})^{-1}_{ij}$ should here be interpreted as the $(i,j)$th element of the inverted matrix $(T_{\bm{v}})^{-1}$. All of the expectations in Eq~\eqref{eq:mu_iv} are means of rectangularly-truncated MNDs.\\

The third step of calculating the mean of a ST-MND is to express the means of hyperrectangularly-truncated MNDs described in Eq~\eqref{eq:mu_iv} in terms of integrals of hyperrectangularly-truncated MNDs. This relationship has been previously derived in \cite{Manjunath2021} via the moment generating function approach, and for the bivariate case with zero mean this relationship is
\begin{equation}
\mathbb{E}(W_i; \bm{0}, \varepsilon, \bm{a}, \bm{b}) = \displaystyle\sum_{k=1}^2 \varepsilon_{ik} \left( F_k(a_k; \bm{0}, \varepsilon, \bm{a}, \bm{b}) - F_k(b_k; \bm{0}, \varepsilon, \bm{a}, \bm{b}) \right),
\label{eq:E_Wi}
\end{equation}
where $\varepsilon_{ik}$ is the $(i,k)$th element of the covariance matrix $\varepsilon$, and $F_k(w_k; \bm{0}, \varepsilon, \bm{a}, \bm{b})$ is the $k$th univariate marginal density for a two-dimensional truncated MND whose probability density is given by Eq~\eqref{eq:phi_R}. Following~\cite{Cartinhour1990}, these univariate marginal densities for a two-dimensional truncated MND can be expressed as
\begin{align}
\label{eq:F1}
F_1(w_1;\bm{0},\varepsilon,\bm{a},\bm{b}) &= \dfrac{ \phi(w_1;0,\varepsilon_{11}) \,\, \Phi \left(a_2,b_2; w_1 \varepsilon_{12}/\varepsilon_{11}, |\varepsilon|/\varepsilon_{11} \right)}{\Phi(\bm{a},\bm{b};\bm{0},\varepsilon)}, \\ 
F_2(w_2;\bm{0},\varepsilon,\bm{a},\bm{b}) &= \dfrac{\phi(w_2; 0, \varepsilon_{22}) \,\, \Phi \left(a_1,b_1; w_2 \varepsilon_{12}/\varepsilon_{22}, |\varepsilon|/\varepsilon_{22} \right)}{ \Phi(\bm{a},\bm{b};\bm{0},\varepsilon)}, 
\label{eq:F2}
\end{align}
where the probability density functions $\phi(\cdot)$ and integrals $\Phi(\cdot)$ on the numerators of these equations are for unbounded and truncated univariate normal distributions respectively, and the integrals on the denominators are for rectangularly-truncated bivariate normal distributions. All of these quantities can be readily computed using modern software.

\subsubsection*{Semi-analytical method: Covariance of the simplex-truncated bivariate normal distribution}
Elements $(\sigma_{\T})_{ij}$ of the covariance matrix for a ST-MND can be calculated using the semi-analytical method in a similar manner to the calculation of mean vector elements $(\mu_{\T})_i$ that was described in the previous section, albeit with slightly more complicated formulae. From Eq~\eqref{eq:elements}, $(\sigma_\T)_{ij} = \mathbb{E}(X_i X_j) - \mathbb{E}(X_i) \mathbb{E}(X_j)$, and the latter expectations $\mathbb{E}(X_i)$ and $\mathbb{E}(X_j)$ for ST-MNDs are equal to mean vector elements $(\mu_{\T})_i$ and $(\mu_{\T})_j$ so can already be calculated as in the previous section. The element $(\mu_\T)_{ij}$ is used to denote the expectation $\mathbb{E}(X_i X_j)$ for a ST-MND. Analogously to Eq~\eqref{eq:mu_T_conditional}, for the bivariate case this expectation is calculated as
\begin{multline}
(\mu_{\T})_{ij} = \dfrac{1}{Z} \left( \mu_{ij} - \Phi_1 \mu_{ij,1} - \Phi_2 \mu_{ij,2} - \Phi_3 \mu_{ij,3} \right. \\ \left. + \, \Phi_{12} \mu_{ij,12} + \Phi_{13} \mu_{ij,13} + \Phi_{23} \mu_{ij,23} \right), \qquad i=1,2,
\label{eq:mu_ij_T_conditional}
\end{multline}
where 
\begin{equation}
\mu_{ij} = \mu_i \mu_j + \sigma_{ij},
\end{equation}
and $\mu_{ij,\bm{v}}$ is the expectation $\mathbb{E}(X_i X_j)$ for the bivariate normal distribution $\mathcal{N}(\bm{\mu},\Sigma)$ truncated within the domain $\mathcal{L}_{\bm{v}}$ (see Table~\ref{table:R}). Expressions for $\Phi_{\bm{v}}$ and $Z$ were already derived in a previous section, and analogously to Eq~\eqref{eq:mu_iv_before} each $\mu_{ij,\bm{v}}$ can be expressed as
\begin{equation}
\mu_{ij,\bm{v}} = \mathbb{E}\left( X_i X_j ; \,\, \bm{0}, \,\, T_{\bm{v}} \Sigma T_{\bm{v}}^\top, \,\, \bm{c}_{\bm{v}} - T_{\bm{v}} \bm{\mu}, \,\, \bm{d}_{\bm{v}} - T_{\bm{v}} \bm{\mu} \right),
\end{equation}
and noting from Eq~\eqref{eq:transform} that $\bm{x} = T_{\bm{v}}^{-1} \bm{w} + \bm{\mu}$,
\begin{align}
\nonumber &\mu_{ij,\bm{v}} = \mu_i \mu_j \\
\nonumber &\quad \quad+ \, \mu_i \displaystyle\sum_{m=1}^2 (T_{\bm{v}})_{jm}^{-1} \, \mathbb{E} \left(W_m; \, \bm{0}, \, T_{\bm{v}} \Sigma T_{\bm{v}}^\top, \, \bm{c}_{\bm{v}} - T_{\bm{v}} \bm{\mu}, \, \bm{d}_{\bm{v}} - T_{\bm{v}} \bm{\mu} \right) \\
\nonumber &\quad \quad+ \, \mu_j \displaystyle\sum_{k=1}^2 (T_{\bm{v}})_{ik}^{-1} \, \mathbb{E} \left(W_k ; \, \bm{0}, \, T_{\bm{v}} \Sigma T_{\bm{v}}^\top, \, \bm{c}_{\bm{v}} - T_{\bm{v}} \bm{\mu}, \, \bm{d}_{\bm{v}} - T_{\bm{v}} \bm{\mu} \right) \\
&\quad \quad+ \, \displaystyle\sum_{k=1}^2 \displaystyle\sum_{m=1}^2 (T_{\bm{v}})_{ik}^{-1} \, (T_{\bm{v}})_{jm}^{-1} \, \mathbb{E} \left(W_k W_m; \, \bm{0}, \, T_{\bm{v}} \Sigma T_{\bm{v}}^\top, \, \bm{c}_{\bm{v}} - T_{\bm{v}} \bm{\mu}, \, \bm{d}_{\bm{v}} - T_{\bm{v}} \bm{\mu} \right).
\label{eq:mu_ijv}
\end{align}
Expectations appearing in the first two summations within Eq~\eqref{eq:mu_ijv} can already be evaluated using Eq~\eqref{eq:E_Wi}-\eqref{eq:F2}. The only remaining expectations in Eq~\eqref{eq:mu_ijv} to evaluate are of the form $\mathbb{E} \left(W_k W_m; \, \bm{0}, \, T_{\bm{v}} \Sigma T_{\bm{v}}^\top, \, \bm{c}_{\bm{v}} - T_{\bm{v}} \bm{\mu}, \, \bm{d}_{\bm{v}} - T_{\bm{v}} \bm{\mu} \right)$, each of which represents the second (raw) moment of a rectangularly-truncated bivariate normal distribution defined in coordinates $\bm{w}$, bounded by $\bm{c}_{\bm{v}} - T_{\bm{v}} \bm{\mu} < \bm{w} < \bm{d}_{\bm{v}} - T_{\bm{v}} \bm{\mu}$, with zero mean and covariance matrix $ T_{\bm{v}} \Sigma T_{\bm{v}}^\top$. The second moment of hyperrectangularly-truncated MNDs has been previously derived in~\cite{Manjunath2021}, and for the bivariate case with zero mean this moment is given by
\begin{align}
\nonumber &\mathbb{E}(W_i W_j; \bm{0},\varepsilon,\bm{a},\bm{b}) = \varepsilon_{ij} \\
\nonumber & \quad \quad + \, \displaystyle\sum_{k=1}^2 \dfrac{\varepsilon_{ik} \varepsilon_{jk}}{\varepsilon_{kk}} \left( a_k F_k(a_k; \bm{0},\varepsilon,\bm{a},\bm{b}) - b_k F_k(b_k; \bm{0}, \varepsilon, \bm{a}, \bm{b}) \right) \\ 
\nonumber & \quad \quad + \, \left[ \dfrac{\varepsilon_{i1}}{\varepsilon_{11}} \left( \varepsilon_{11} \varepsilon_{j2} - \varepsilon_{12} \varepsilon_{j1} \right) + \dfrac{\varepsilon_{i2}}{\varepsilon_{22}} \left( \varepsilon_{22} \varepsilon_{j1} - \varepsilon_{12} \varepsilon_{j2} \right) \right] \\
\nonumber & \quad \quad \times \Bigl( F_{1,2} (a_1,a_2; \bm{0},\varepsilon,\bm{a},\bm{b}) + F_{1,2}(b_1,b_2; \bm{0},\varepsilon,\bm{a},\bm{b}) \\
 & \quad \quad - \, F_{1,2}(a_1,b_2; \bm{0},\varepsilon,\bm{a},\bm{b}) -F_{1,2}(b_1,a_2; \bm{0},\varepsilon,\bm{a},\bm{b}) \Bigr),
\label{eq:E_WiWj}
\end{align}
where $F_k(w_k; \bm{0},\varepsilon,\bm{a},\bm{b})$ is calculated via Eq~\eqref{eq:F1} for $k=1$ and via Eq~\eqref{eq:F2} for $k=2$, and $F_{k,q}(w_k,w_q; \bm{0},\varepsilon, \bm{a}, \bm{b})$ is the $(k,q)$th bivariate marginal density for a two-dimensional truncated MND whose probability density is given in Eq~\eqref{eq:phi_R}. Eq~\eqref{eq:E_WiWj} was obtained from Eq~(16) of \cite{Manjunath2021} by simplifying for the bivariate case and by noticing that the bivariate marginal density function is symmetric with respect to switching indices $k$ and $q$. This latter observation allows all bivariate marginal densities appearing in Eq~\eqref{eq:E_WiWj} to be written with common indices $k=1$ and $q=2$. These bivariate marginal densities can be obtained from readily computed quantities via \cite{Manjunath2021}
\begin{equation}
F_{1,2}(w_1,w_2; \bm{0}, \varepsilon, \bm{a}, \bm{b}) = \dfrac{\phi \left( \bm{w}, \bm{0}, \varepsilon \right)}{\Phi(\bm{a},\bm{b};\bm{0},\varepsilon)}.
\label{eq:F_12}
\end{equation}

\subsubsection*{Semi-analytical method for the simplex-truncated multivariate normal distribution}

Formulae used to obtain the integral $Z$, mean vector $\bm{\mu}_\T$ and covariance matrix $\Sigma_\T$ of a simplex-truncated bivariate normal distribution (i.e.\ ST-MNDs of dimension $n=2$) derived in the previous sections, can be generalised for calculation of $Z$, $\bm{\mu}_\T$ and $\Sigma_\T$ for ST-MNDs of any dimension $n$, using six modifications. In the present section, these modifications are described and justified.

The first modification is to generalise the formulae given in Eq~\eqref{eq:Z_expression}, \eqref{eq:mu_T_conditional} and \eqref{eq:mu_ij_T_conditional}. These formulae use the inclusion-exclusion principle to express $Z$, $\bm{\mu}_\T$ and $\Sigma_\T$ in terms of integrals, mean vector elements and covariance matrix elements of MNDs truncated by a number of constraints equal to or less than their dimension. It will be convenient here to write all covariance matrix elements $(\sigma_\T)_{ij}$ of the ST-MND in terms of mean vector elements $(\mu_\T)_i$ and second (raw) moment elements $(\mu_\T)_{ij}$ via $(\sigma_\T)_{ij} = (\mu_\T)_{ij} - (\mu_\T)_i (\mu_\T)_j$, $\forall i,j=1,...,n$. Then, Eq~\eqref{eq:Z_expression}, \eqref{eq:mu_T_conditional} and \eqref{eq:mu_ij_T_conditional} to generate $Z$, $(\mu_\T)_i$ and $(\mu_\T)_{ij}$ are generalised to the $n$-dimensional case as follows:
\begin{align} \nonumber
&Z = 1 - \displaystyle\sum_{v=1}^{n+1} \phi_v + \displaystyle\sum_{v_1=1}^{n+1} \, \displaystyle\sum_{v_2=v_1+1}^{n+1} \phi_{[v_1\,v_2]} - \displaystyle\sum_{v_1=1}^{n+1} \,\displaystyle\sum_{v_2=v_1+1}^{n+1} \, \displaystyle\sum_{v_3=v_2+1}^{n+1} \phi_{[v_1\,v_2\,v_3]} + ... \\
& \qquad \qquad ... + (-1)^n \underbrace{\displaystyle\sum_{v_1=1}^{n+1} \,\, ... \!\! \displaystyle\sum_{v_n = v_{n-1} + 1}^{n+1}}_{\mbox{$n$ summations}} \phi_{[v_1\,v_2\,...\,v_n]},  \label{eq:Z_general} \\
&\nonumber (\mu_\T)_i = \dfrac{1}{Z} \, \Biggl( \mu_i - \displaystyle\sum_{v=1}^{n+1} \phi_v \mu_{i,v} + \displaystyle\sum_{v_1=1}^{n+1} \, \displaystyle\sum_{v_2=v_1+1}^{n+1} \phi_{[v_1\,v_2]} \, \mu_{i,[v_1\,v_2]}\\
&\qquad \qquad  - \displaystyle\sum_{v_1=1}^{n+1} \,\displaystyle\sum_{v_2=v_1+1}^{n+1} \, \displaystyle\sum_{v_3=v_2+1}^{n+1} \phi_{[v_1\,v_2\,v_3]} \, \mu_{i,[v_1\,v_2\,v_3]} + ... \nonumber \\
&\qquad \qquad \qquad \qquad ... + (-1)^n \underbrace{\displaystyle\sum_{v_1=1}^{n+1} \,\, ... \!\! \displaystyle\sum_{v_n = v_{n-1} + 1}^{n+1}}_{\mbox{$n$ summations}} \phi_{[v_1\,v_2\,...\,v_n]} \, \mu_{i,[v_1\,v_2\,...\,v_n]} \Biggr), \label{eq:mu_i_general}\\
&\nonumber (\mu_\T)_{ij} = \dfrac{1}{Z} \, \Biggl( \mu_{ij} - \displaystyle\sum_{v=1}^{n+1} \phi_v \mu_{ij,v} + \displaystyle\sum_{v_1=1}^{n+1} \, \displaystyle\sum_{v_2=v_1+1}^{n+1} \phi_{[v_1\,v_2]} \, \mu_{ij,[v_1\,v_2]} \\
&\qquad \qquad  - \displaystyle\sum_{v_1=1}^{n+1} \,\displaystyle\sum_{v_2=v_1+1}^{n+1} \, \displaystyle\sum_{v_3=v_2+1}^{n+1} \phi_{[v_1\,v_2\,v_3]} \, \mu_{ij,[v_1\,v_2\,v_3]} + ... \nonumber \\
&\qquad \qquad \qquad \qquad ... + (-1)^n \underbrace{\displaystyle\sum_{v_1=1}^{n+1} \,\, ... \!\! \displaystyle\sum_{v_n = v_{n-1} + 1}^{n+1}}_{\mbox{$n$ summations}} \phi_{[v_1\,v_2\,...\,v_n]} \, \mu_{ij,[v_1\,v_2\,...\,v_n]} \Biggr). \label{eq:mu_ij_general}
\end{align}
To calculate the quantities that appear in Eq~\eqref{eq:Z_general}-\eqref{eq:mu_ij_general}, note that the expression of the integral $\phi_{\bm{v}}$ provided in Eq~\eqref{eq:phi_v} is already applicable to the $n$-dimensional case. Generalisation of expressions for the mean vector elements $\mu_{i,\bm{v}}$ and second (raw) moment elements $\mu_{ij,\bm{v}}$ in any dimension $n$, requires the second modification: the upper limits of the summations in Eq~\eqref{eq:mu_iv} and Eq~\eqref{eq:mu_ijv} are changed from 2 to $n$.

The third modification is the generalisation of the forms of the transformation matrix $T_{\bm{v}}$ and constraint vectors $\bm{c}_{\bm{v}}$, $\bm{d}_{\bm{v}}$ to any dimension $n$, for which the bivariate case is shown in Table~\ref{table:R}. These quantities are required for calculation of $\phi_{\bm{v}}$, $\mu_{i,\bm{v}}$ and $\mu_{ij,\bm{v}}$. For a domain $\mathcal{L}_{\bm{v}}$ defined by a vector $\bm{v} = [v_1\,v_2\,v_3\,...\,v_q]$ containing $q$ elements,  $T_{\bm{v}}$, $\bm{c}_{\bm{v}}$ and $\bm{d}_{\bm{v}}$ can be obtained via the following two-step algorithm:
\begin{description}
\item[Step 1] Initialise the transformation matrix and constraint vectors so that they represent no spatial restrictions on the $n$-dimensional space. (All constraints are added in the second step of the algorithm.) This is achieved by setting the transformation matrix to be equal to the $n$-dimensional identity matrix $I_n$, setting the lower constraint vector $\bm{c}_{\bm{v}}$ to consist of $n$ elements all of which are negative infinity, and setting the upper constraint vector $\bm{d}_{\bm{v}}$ to consist of $n$ elements all of which are positive infinity:
\[
T_{\bm{v}} = I_n, \qquad \bm{c}_{\bm{v}} = [ -\infty \,\,\,\, ... \,\,\,\, -\infty]^\top, \qquad \bm{d}_{\bm{v}} = [ \infty \,\,\,\, ... \,\,\,\, \infty]^\top.
\]
\item[Step 2] For all elements $i=1,...,q$ of $\bm{v} = [v_1\,v_2\,v_3\,...\,v_q]$:

If $v_i \leq n$, replace the $i$th element of $\bm{d}_{\bm{v}}$ with zero, i.e.\ $(\bm{d}_{\bm{v}})_i \leftarrow 0$. This element replacement is equivalent to introducing the restriction $x_i < 0$ that ``slices off'' the region of space associated with the ST-MND truncation $x_i \geq 0$.

Otherwise, $v_i=n+1$, so identify any one row $j$ of $T_{\bm{v}}$ where $j \notin \bm{v}$. For this value of $j$, replace all elements in the $j$th row of $T_{\bm{v}}$ with one, i.e.\ $(T_{\bm{v}})_{jk} \leftarrow 1, \forall k=1,...,n$, and replace the $j$th element of $\bm{c}_{\bm{v}}$ with one, i.e.\ $(\bm{c}_{\bm{v}})_j \leftarrow 1$. These element replacements are together equivalent to introducing the restriction $\displaystyle\sum_{i=1}^n x_i > 1$ that ``slices off'' the region of space associated with the ST-MND truncation~$\displaystyle\sum_{i=1}^n x_i \leq 1$.
\end{description}
For the bivariate case, this two-step algorithm yields the transformation matrices $T_{\bm{v}}$ and constraint vectors $\bm{c}_{\bm{v}}$ and $\bm{d}_{\bm{v}}$ shown in Table~\ref{table:R}.\\

The fourth modification is to generalise the expressions for the first and second (raw) moments of the hyperrectangularly-truncated MNDs possessing zero mean that require calculation in the generalised forms of Eq~\eqref{eq:mu_iv} and Eq~\eqref{eq:mu_ijv}. For the first raw moment $\mathbb{E}(W_i; \bm{0}, \varepsilon, \bm{a}, \bm{b})$, Eq~\eqref{eq:E_Wi} can be used for the $n$-dimensional case by changing the upper limit on the summation present in this equation from 2 to $n$.  The $n$-dimensional generalisation of the second raw moment $\mathbb{E}(W_i W_j; \bm{0}, \varepsilon, \bm{a}, \bm{b})$ requires replacing Eq~\eqref{eq:E_WiWj} by
\begin{align}
& \nonumber \mathbb{E}(W_iW_j; \bm{0},\varepsilon, \, \bm{a},\bm{b}) = \varepsilon_{ij} \\
& \quad \quad \nonumber  + \, \displaystyle\sum_{k=1}^n \, \dfrac{\varepsilon_{ik} \varepsilon_{jk}}{\varepsilon_{kk}} \left( a_k F_k(a_k; \bm{0},\varepsilon,\bm{a},\bm{b}) - b_k F_k(b_k; \bm{0}, \varepsilon, \bm{a}, \bm{b}) \right) \\ 
& \quad \quad \nonumber  + \, \displaystyle\sum_{k=1}^n \, \displaystyle\sum_{q=k+1}^n \, \left[ \dfrac{\varepsilon_{ik}}{\varepsilon_{kk}} \left( \varepsilon_{kk} \varepsilon_{jq} - \varepsilon_{kq} \varepsilon_{jk} \right) + \dfrac{\varepsilon_{iq}}{\varepsilon_{qq}} \left(\varepsilon_{qq} \varepsilon_{jk} - \varepsilon_{qk} \varepsilon_{jq} \right) \right] \\
& \quad \quad \nonumber \times \Bigl( F_{k,q} (a_k,a_q; \bm{0},\varepsilon,\bm{a},\bm{b}) + F_{k,q}(b_k,b_q; \bm{0},\varepsilon,\bm{a},\bm{b}) \\
 & \quad \quad - \, F_{k,q}(a_k,b_q; \bm{0},\varepsilon,\bm{a},\bm{b}) -F_{k,q}(b_k,a_q; \bm{0},\varepsilon,\bm{a},\bm{b}) \Bigr).
 \label{eq:E_WiWj_general}
\end{align}
This equation was obtained by combining Eq~(16) of~\cite{Manjunath2021} with the observation that the bivariate marginal density $F_{k,q}(w_k,w_q;\bm{0},\varepsilon,\bm{a},\bm{b})$ is symmetric with respect to switching indices $k$ and $q$.\\

The fifth modification is to generalise the expression for the univariate marginal density function of a hyperrectangularly-truncated MND possessing zero mean, $F_k(w_k;\bm{0},\varepsilon,\bm{a},\bm{b})$, to the $n$-dimensional case, as this marginal density function requires evaluation to obtain the first and second (raw) moments discussed in the fourth modification. Following \cite{Cartinhour1990}, this generalisation requires replacing Eq~\eqref{eq:F1} and \eqref{eq:F2} by
\begin{equation}
F_k(w_k; \bm{0}, \varepsilon, \bm{a}, \bm{b}) = \dfrac{\phi \left(w_k;0,\varepsilon_{kk}\right) \Phi \left(\bm{a}_{(-k)},\bm{b}_{(-k)}; \dfrac{w_k \varepsilon_{(-k,+k)}}{\varepsilon_{kk}}, \left[ \varepsilon_{(-k,-k)}^{-1} \right]^{-1} \right)}{\Phi(\bm{a},\bm{b}; \bm{0},\varepsilon)},
\label{eq:F_general}
\end{equation}
where $\bm{a}_{(-k)}$  and $\bm{b}_{(-k)}$ should be understood as the vectors $\bm{a}$ and $\bm{b}$ with their $k$th elements removed, $\varepsilon_{(-k,+k)}$ is a column vector consisting of the $k$th column of $\varepsilon$ with its $k$th row removed, i.e.\ $\varepsilon_{(-k,+k)} = [\varepsilon_{1k}\,\,\,\,\varepsilon_{2k} \,\,\,\, ... \,\,\,\,\varepsilon_{(k-1)k} \,\,\,\, \varepsilon_{(k+1)k} \,\,\,\, ... \,\,\,\, \varepsilon_{nk}]^\top$, and $\varepsilon_{(-k,-k)}^{-1}$ is obtained by firstly finding the matrix inversion of $\varepsilon$ and secondly removing the $k$th column and $k$th row of the resulting inverse matrix. The numerator of Eq~\eqref{eq:F_general} is the product of a probability density function for a univariate normal distribution and the integral of a hyperrectangularly-truncated MND of dimension ($n-1$), and the denominator of Eq~\eqref{eq:F_general} is the integral of a hyperrectangularly-truncated MND of dimension $n$.\\

The sixth and final modification is to generalise the expression for the bivariate marginal density function of a hyperrectangularly-truncated MND possessing zero mean, $F_{k,q}(w_k,w_q;\bm{0},\varepsilon,\bm{a},\bm{b})$, to the $n$-dimensional case, as this marginal density function requires evalution to obtain the second (raw) moment shown in Eq~\eqref{eq:E_WiWj_general}. After combining Eq~(21)-(23) of~\cite{Manjunath2021} with some simplifications that remove a distribution transform, the required bivariate marginal density function can be expressed, for any $k \neq q$, as
\begin{multline}
\quad F_{k,q}(w_k,w_q; \bm{0}, \varepsilon, \bm{a}, \bm{b}) = \phi \left( \left[ w_k \,\, w_q \right]^\top; \bm{0}, \varepsilon_{(+k+q,+k+q)} \right) \\ \times \, \Phi \left((\bm{a}_{kq}^*)_{(-k-q)}, (\bm{b}_{kq}^*)_{(-k-q)}; \bm{0}, (R_{kq})_{(-k-q),(-k-q)} \right) \, / \, \, \Phi(\bm{a},\bm{b}; \bm{0},\varepsilon), \quad
\label{eq:F_kq}
\end{multline}
where $\varepsilon_{(+k+q,+k+q)}$ should be understood as the 2$\times$2 matrix consisting of elements belonging only to the $k$th and $q$th rows, and $k$th and $q$th columns, of the covariance matrix $\varepsilon$,
\begin{equation}
\varepsilon_{(+k+q,+k+q)} = \begin{bmatrix} \varepsilon_{kk} & \varepsilon_{kq} \\ \varepsilon_{qk} & \varepsilon_{qq} \end{bmatrix},
\end{equation}
$(R_{kq})_{(-k-q),(-k-q)}$ should be understood as the matrix of second-order partial correlation coefficients, $R_{kq}$, with its $k$th row, $q$th row, $k$th column and $q$th column removed, and the $(n-2)$-dimensional column vectors $(\bm{a}_{kq}^*)_{(-k-q)}$ and $(\bm{b}_{kq}^*)_{(-k-q)}$ are obtained from corresponding $n$-dimensional column vectors $\bm{a}_{kq}^*$ and $\bm{b}_{kq}^*$, via removal of both their $k$th and $q$th elements. Elements $(a_{kq}^*)_i$ and $(b_{kq}^*)_i$ of these column vectors $\bm{a}_{kq}^*$ and $\bm{b}_{kq}^*$ are calculated as
\begin{align}
&(a_{kq}^*)_i = \dfrac{a_i/\sqrt{\varepsilon_{ii}} - \beta_{ik,q} w_k/\sqrt{\varepsilon_{kk}} - \beta_{iq,k} w_q/\sqrt{\varepsilon_{qq}} }{ \sqrt{ \left(1 - \rho_{iq}^2 \right) \left(1 - \rho_{ik,q}^2 \right)} }, \qquad  i=1,..,n,\\
&(b_{kq}^*)_i = \dfrac{b_i/\sqrt{\varepsilon_{ii}} - \beta_{ik,q}  w_k/\sqrt{\varepsilon_{kk}} - \beta_{iq,k} w_q/\sqrt{\varepsilon_{qq}} }{ \sqrt{ \left(1 - \rho_{iq}^2 \right) \left(1 - \rho_{ik,q}^2 \right)} }, \qquad  i=1,...,n,
\label{eq:b_i_star}
\end{align}
where $\rho_{ij,m}$, $\beta_{ij,m}$, and $\rho_{ij}$ are first-order partial correlation coefficients, partial regression coefficients, and bivariate correlation coefficients, respectively. Denoting $\rho_{ij,kq}$ as the $(i,j)$th element of the matrix $R_{kq}$, the elements $\rho_{ij,kq}$, $\rho_{ij,m}$, $\beta_{ij,m}$ and $\rho_{ij}$ can all be calculated from elements $\varepsilon_{ij}$ of the covariance matrix $\varepsilon$ via general formulae that are included here for completeness:
\begin{align}
&\rho_{ij,kq} = \left( \rho_{ij,k} - \rho_{iq,k} \rho_{jq,k} \right) / \sqrt{ \left( 1 - \rho_{iq,k}^2 \right) \left( 1 - \rho_{jq,k} ^2 \right)},
\label{eq:rho_ijkq} \\
&\rho_{ij,m} = \left(\rho_{ij}-\rho_{im} \rho_{jm} \right) / \sqrt{ \left(1 - \rho_{im}^2\right) \left(1 - \rho_{jm}^2 \right)},
\label{eq:rho_ijk} \\
&\beta_{ij,m} = \left( \rho_{ij}- \rho_{im}\rho_{jm} \right) / \left(1 - \rho_{jm}^2 \right),
\label{eq:beta_ijk} \\
&\rho_{ij} = \dfrac{\varepsilon_{ij}}{\sqrt{\varepsilon_{ii} \varepsilon_{jj}}}.
\label{eq:rho_ij}
\end{align}

\subsection*{Comparing the three methods}
To explore the relative usefulness of each of the three methods described here (naive rejection sampling, the Gessner \textit{et al}.\ method, and the semi-analytical method), in the present work these methods were compared both for their accuracy and their speed, to simultaneously estimate the integral, mean and covariance of ST-MNDs with dimensions $n=2$ (i.e.\ bivariate) or higher. To perform this comparison, parameters $\bm{\mu}$ and $\Sigma$ were sampled for 100 ST-MNDs per dimension, and the speed and accuracy for each of the three methods, to estimate the quantities $Z$, $\bm{\mu}_{\T}$ and $\Sigma_\T$ defined in Eq~\eqref{eq:Z} and \eqref{eq:elements}, was assessed.

For each of the 100 distributions per dimension, sampling of the parameters $\bm{\mu}$ and $\Sigma$ proceeded as follows. First, the mean vector $\bm{\mu}$ was sampled from a uniform distribution within the unit simplex, as ST-MNDs used to represent compositional data will likely (but not necessarily) possess mean vectors $\bm{\mu}$ satisfying this criterion. To accomplish this sampling in practice, for a distribution of dimension $n$, each of the $n$ elements $\mu_i$ of the mean vector $\bm{\mu}$ was randomly sampled from an uniform distribution bounded between zero and one, $\mu_i \sim \mathcal{U}(0,1), \, \forall i=1,...,n$. If the sum of all $n$ elements in the sample $\bm{\mu}$ was less than or equal to one, $\displaystyle\sum_{i=1}^n \mu_i \leq 1$, the mean vector was kept for subsequent comparison of the three methods. Otherwise, the sample $\bm{\mu}$ was discarded, and all elements of $\bm{\mu}$ were resampled from $\mathcal{U}(0,1)$. This process continued until a sampled  $\bm{\mu}$ satisfied $\displaystyle\sum_{i=1}^n \mu_i \leq 1$.

Second, the sampling procedure for the covariance matrix $\Sigma$ aimed to balance plausibility of the resulting distributions for modelling compositional data and practical considerations. All of the diagonal elements of $\Sigma$ were sampled from a uniform distribution, $\sigma_{ii} \sim \mathcal{U}(0,0.25), \, \forall i=1,...,n$, as this choice represents plausible uncertainty in compositional data. Then, all off-diagonal elements of $\Sigma$ were obtained by sampling correlation coefficients $\rho_{ij} \sim \mathcal{U}(-0.5,0.5), \, \forall i=1,...,n, \, j=i+1,..,n,$ and thereafter calculating $\sigma_{ij} = \sigma_{ji} = \rho_{ij} \sqrt{ \sigma_{ii} \sigma_{jj} }$. Since covariance matrices must be positive semi-definite, and the sampling procedure described so far guarantees that the sampled matrix $\Sigma$ is real and symmetric so that its eigenvalues will be real, all that is left is to check if all eigenvalues of $\Sigma$ are non-negative for this matrix to be positive semi-definite and thus satisfy the definition of a covariance matrix. Thus, the sampled matrix $\Sigma$ was kept only if all of its eigenvalues were non-negative. Otherwise, this matrix was discarded, and all of its elements (diagonal and non-diagonal) were resampled according to the previously described procedure. This process continued until a sampled matrix $\Sigma$ possessed only non-negative eigenvalues. Sampling correlation coefficients $\rho_{ij} \sim \mathcal{U}(-0.5,0.5)$ ensured that covariance matrices $\Sigma$ could be found in a reasonable computational time. (In practice, it was found that if instead the full possible range of values for correlation coefficients $-1 \leq \rho_{ij} \leq 1$ was used for sampling non-diagonal elements of $\Sigma$, the ratio of accepted to rejected matrices became impractically small for distributions of dimension $n=9$ or higher.)

Implementation of all three methods was carried out in MATLAB R2021b. For the naive rejection sampling method, $10^4$ samples were used for estimating the integral, mean and covariance of each ST-MND. Analogously, for the Gessner \textit{et al}.\ method, $10^4$ samples were used for the unbiased estimate of the integral (in the Holmes-Diaconis-Ross algorithm), and for the estimates of the mean and covariance. For reasons of computational practicality (see Results for more details), the naive rejection sampling method was tested for ST-MNDs of dimensions $n=2$ up to $n=7$, the Gessner \textit{et al}.\ method was tested for ST-MNDs of dimensions $n=2$ up to $n=10$, and the semi-analytical method was tested for ST-MNDs of dimensions $n=2$ up to $n=5$.

In the semi-analytical method, computation of integrals of hyperrectangularly-truncated MNDs were carried out using MATLAB's ``mvncdf'' function with default options. However, the accuracy of the semi-analytical method became substantially reduced for ST-MNDs of dimension $n=5$ (see Results), so the possibility that inaccuracies in the computation of integrals of hyperrectangularly-truncated MNDs may be responsible for this observation was explored. Thus, for ST-MNDs of dimension $n=5$, the semi-analytical method was applied a second time to each of the 100 tested ST-MNDs. Specifically, in this additional usage of the semi-analytical method, the maximum absolute error tolerance in MATLAB's ``mvncdf'' function was temporarily changed from $10^{-4}$ (default value when $n \geq 4$) to $10^{-6}$. This temporary change increases the accuracy of computation of integrals of hyperrectangularly-truncated MNDs, that are required within the semi-analytical method.

For the Gessner \textit{et al}.\ method, thinning of the samples was carried out following the settings recommended in~\cite{Gessner2020}: keeping every tenth sample (i.e.\ thinning ratio of ten) for subset simulation, but keeping every second sample (i.e.\ thinning ratio of two) for the algorithms used to directly obtain unbiased estimates of the integral, mean and covariance of ST-MNDs. However, discrepancies in the predictions between the Gessner \textit{et al}.\ and naive rejection sampling methods became apparent for ST-MNDs of dimension $n=7$ (see Results), so the possibility that the correlation of samples used in the Gessner \textit{et al}.\ method may be responsible for this observation was explored. Thus, for ST-MNDs of dimension $n=7$, the Gessner \textit{et al}.\ method was applied a second time to each of the 100 tested ST-MNDs. Specifically, in this additional usage of the Gessner \textit{et al}.\ method, the thinning ratio was temporarily set to ten for \textit{all} algorithms within this method, and its predictions were compared to those obtained using the settings recommended in~\cite{Gessner2020}. This temporary change reduces the correlation of samples used in the Gessner \textit{et al}.\ method.

The implementation of all three methods in MATLAB used here, including all output files generated for the results of this manuscript, are available in the ancillary files.

\section*{Results}
\begin{figure}[!h]
\includegraphics[width=\textwidth]{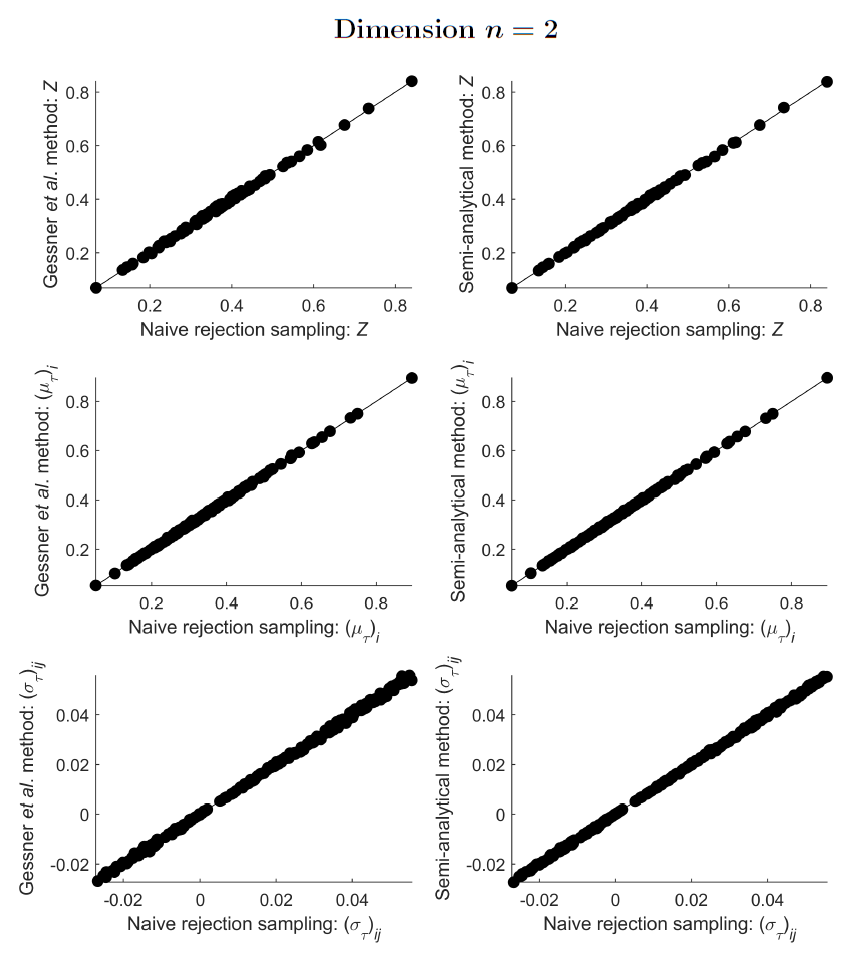}
\caption{{\bf Comparison of estimates of $Z$, and estimates of the elements of $(\mu_\T)_i$ and $(\sigma_\T)_{ij}$, obtained from implementation of the three methods presented in this paper, for 100 different simplex-truncated bivariate normal distributions $\phi_{\T}(\bm{x}; \bm{\mu},\Sigma)$.}}
\label{fig:results_dimension_2}
\end{figure}

\begin{figure}[!h]
\includegraphics[width=\textwidth]{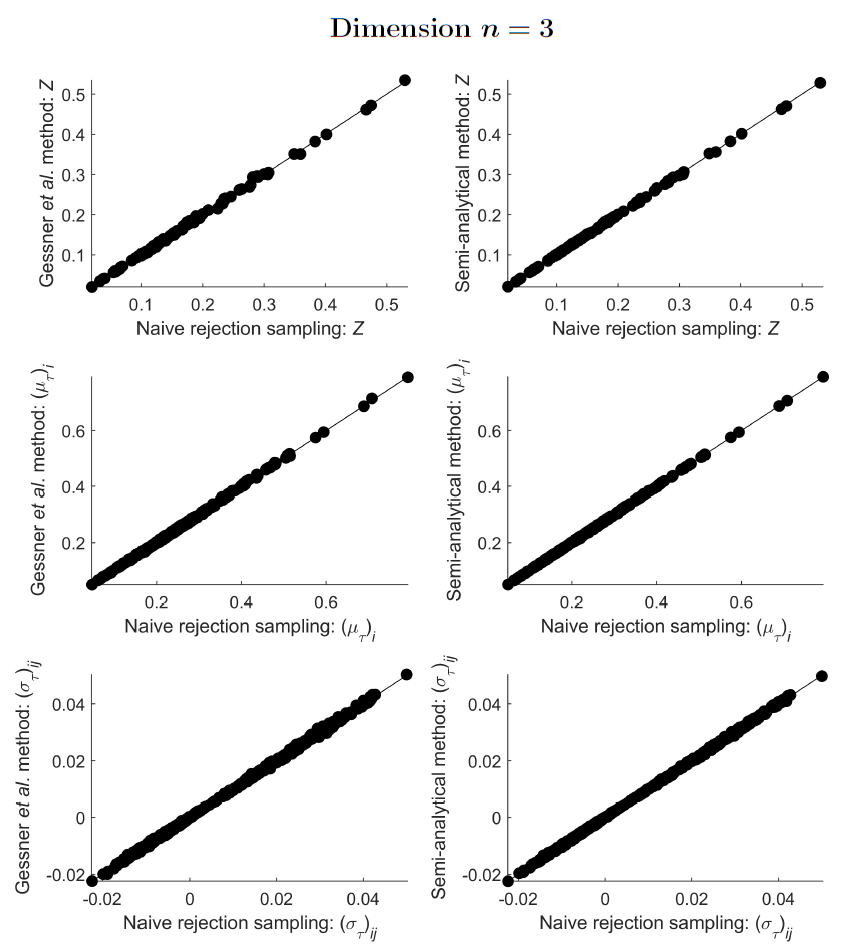}
\caption{{\bf Comparison of estimates of $Z$, and estimates of the elements of $(\mu_\T)_i$ and $(\sigma_\T)_{ij}$, obtained from implementation of the three methods presented in this paper, for 100 different simplex-truncated trivariate normal distributions $\phi_{\T}(\bm{x}; \bm{\mu},\Sigma)$.}}
\label{fig:results_dimension_3}
\end{figure}

\begin{figure}[!h]
\includegraphics[width=\textwidth]{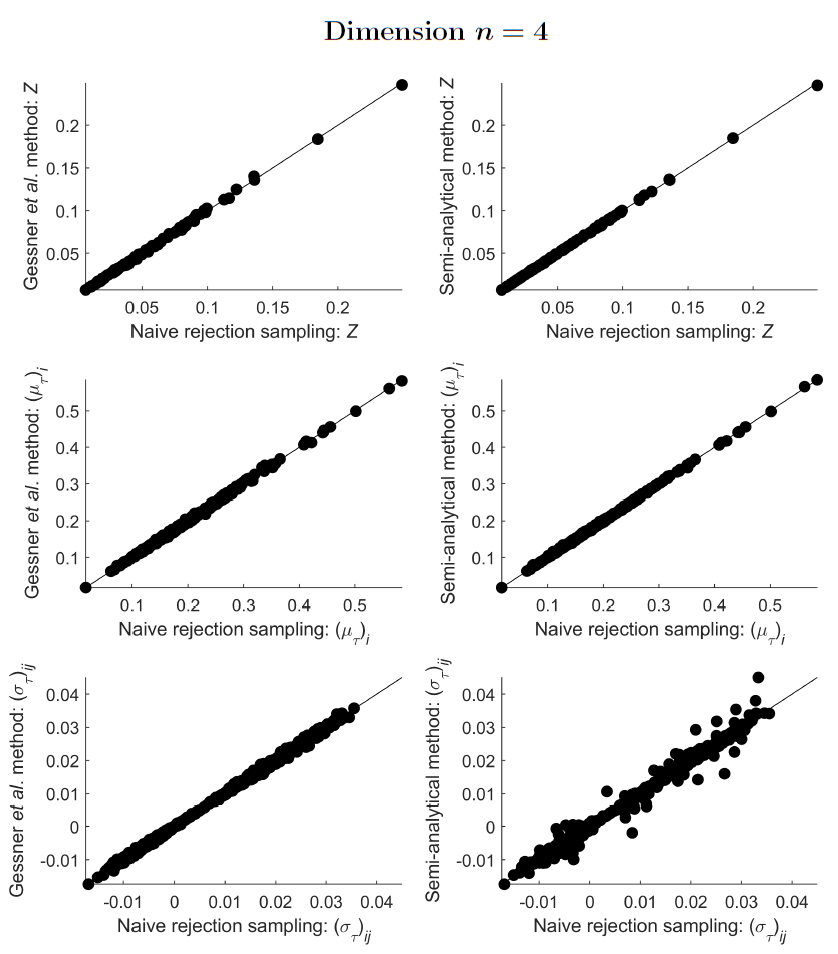}
\caption{{\bf Comparison of estimates of $Z$, and estimates of the elements of $(\mu_\T)_i$ and $(\sigma_\T)_{ij}$, obtained from implementation of the three methods presented in this paper, for 100 different simplex-truncated multivariate normal distributions $\phi_{\T}(\bm{x}; \bm{\mu},\Sigma)$ of dimension $n=4$.} Notice that the semi-analytical method (plots in the right column) is starting to become inaccurate.}
\label{fig:results_dimension_4}
\end{figure}

All three tested methods (naive rejection sampling, the Gessner \textit{et al}.\ method, and the semi-analytical method) agreed well with each other in their estimates of the integral, mean and covariance for ST-MNDs of dimensions $n=2$ and $n=3$ (Fig~\ref{fig:results_dimension_2} and \ref{fig:results_dimension_3}). For ST-MNDs of dimension $n=4$, the two sampling methods (naive rejection sampling and the Gessner \textit{et al}.\ method) compared favourably (plots in the left column of Fig~\ref{fig:results_dimension_4}), but the semi-analytical method produced slightly different estimates of covariance matrix elements (bottom-right plot of Fig~\ref{fig:results_dimension_4}). A plausible explanation for this observation is that the two sampling methods remain accurate but the semi-analytical method loses accuracy when $n \geq 4$ because, in the presently used implementation of the semi-analytical method, the underlying algorithm that computes integrals of hyperrectangularly-truncated MNDs (``mvncdf'' function in MATLAB R2021b) differs between dimensions $n < 4$ and $n \geq 4$~\cite{MathWorks2022}.

\begin{figure}[!h]
\includegraphics[width=\textwidth]{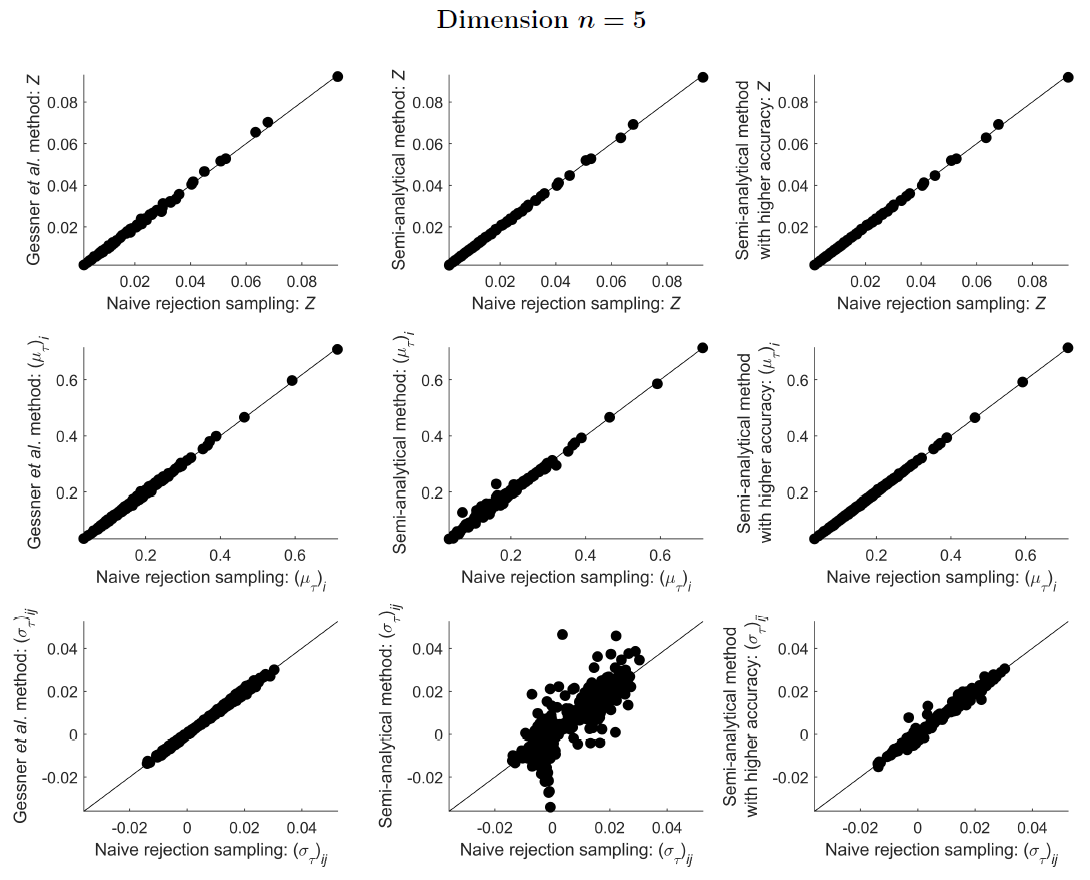}
\caption{{\bf Comparison of estimates of $Z$, and estimates of the elements of $(\mu_\T)_i$ and $(\sigma_\T)_{ij}$, obtained from implementation of the three methods presented in this paper, for 100 different simplex-truncated multivariate normal distributions $\phi_{\T}(\bm{x}; \bm{\mu},\Sigma)$ of dimension $n=5$.} Notice that the semi-analytical method (plots in the central column) becomes rather inaccurate, but increasing the accuracy of the integrals of hyperrectangularly-truncated MNDs calculated within this method vastly improves the estimates overall (plots in the right-most column). However, for one of the distributions tested, the covariance matrix was very poorly estimated by the semi-analytical method with increased accuracy (bottom-right plot, results not shown), with the values of some covariance matrix elements predicted to be more than ten orders of magnitude higher than their values predicted by the other methods.}
\label{fig:results_dimension_5}
\end{figure}

For ST-MNDs of dimension $n=5$, the two sampling methods again compared favourably (plots in the left column of Fig~\ref{fig:results_dimension_5}), but the semi-analytical method yielded substantially different estimates for covariance matrix elements (central bottom plot of Fig~\ref{fig:results_dimension_5}). To investigate this issue further, the semi-analytical method was used a second time for calculations on ST-MNDs of dimension $n=5$, but with an increased accuracy for the computation of integrals of hyperrectangularly-truncated MNDs required within this method (see ``Materials and methods'' for details). Overall, this change substantially improved the correspondence between the semi-analytical method and the naive rejection sampling method (plots in the right column of Fig~\ref{fig:results_dimension_5}). This strongly suggests that the present implementation of the semi-analytical method requires increased accuracy in its computation of integrals of hyperrectangularly-truncated MNDs to obtain accurate estimates of the integral, mean and covariance of ST-MNDs for dimension $n=5$.

However, increasing the accuracy for the computation of integrals of hyperrectangularly-truncated MNDs did not guarantee that the predictions of the semi-analytical method were improved. For one of the 100 ST-MNDs of dimension $n=5$ that was tested, the covariance matrix was so poorly estimated by the semi-analytical method with increased accuracy that some of the covariance matrix elements were more than ten orders of magnitude higher than their values predicted by other methods (see caption of Fig~\ref{fig:results_dimension_5}). Thus, for the semi-analytical method, it is likely that the software package's ability to accurately estimate integrals of hyperrectangularly-truncated MNDs is the limiting factor on the accuracy of this method to estimate the integral, mean and covariance of ST-MNDs.

\begin{figure}[!h]
\includegraphics[width=\textwidth]{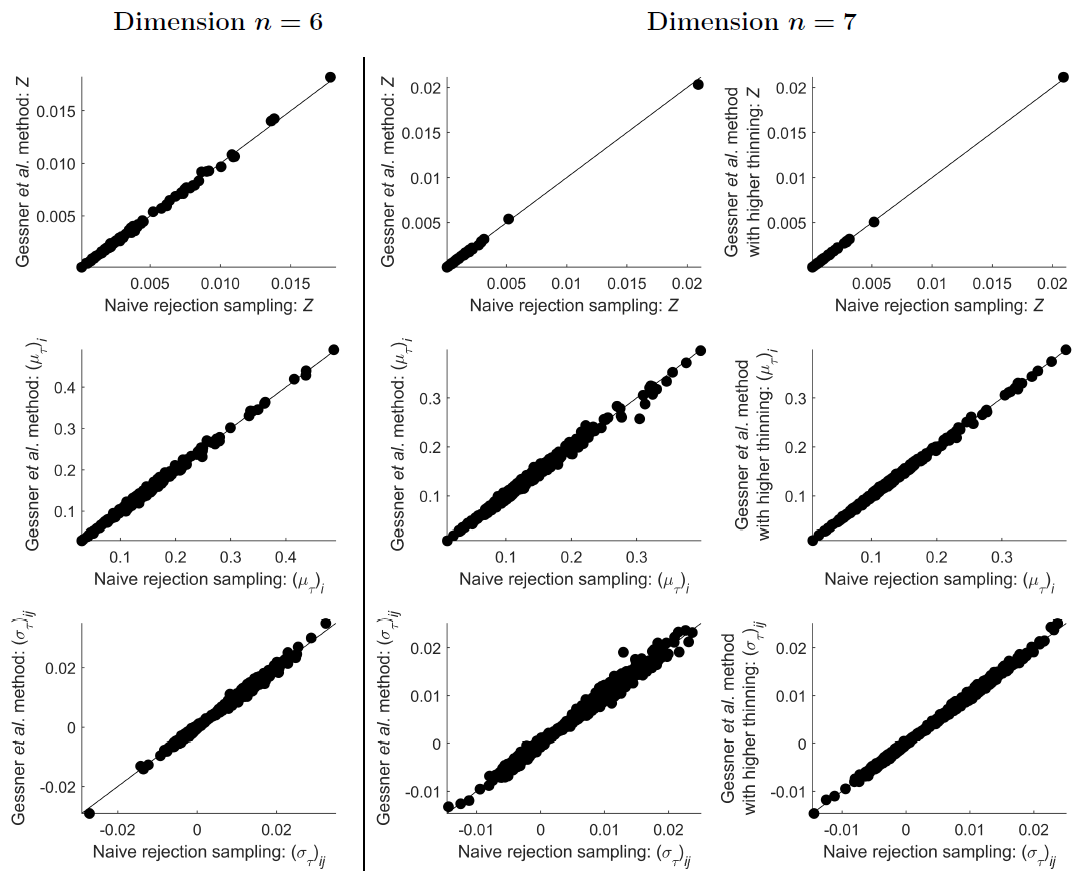}
\caption{{\bf Comparison of estimates of $Z$, and estimates of the elements of $(\mu_\T)_i$ and $(\sigma_\T)_{ij}$, obtained from implementation of the two sampling methods presented in this paper, for 100 different simplex-truncated multivariate normal distributions $\phi_{\T}(\bm{x}; \bm{\mu},\Sigma)$ of dimension $n=6$ (plots in the left column), $n=7$ (plots in the central column), and $n=7$ for a higher thinning ratio used within the Gessner \textit{et al.} method (plots in the right column).} Notice that increasing the thinning ratio in the Gessner \textit{et al.} method improves the match of the two sampling methods (compare plots in the central and right columns).}
\label{fig:results_dimensions_6_and_7}
\end{figure}

For ST-MNDs of dimensions $n=6$ and $n=7$, the two sampling methods continued to agree well (Fig~\ref{fig:results_dimensions_6_and_7}), although slight differences were seen in their estimates of covariance matrix elements for $n=7$ (bottom plot in the central column of Fig~\ref{fig:results_dimensions_6_and_7}). The possibility that the correlation of samples in the Gessner \textit{et al}.\ method may responsible for these slight differences was investigated, by using the Gessner \textit{et al}.\ method a second time, with a higher thinning ratio for all algorithms used to unbiasedly estimate the integral, mean and covariance of ST-MNDs within this method (see ``Materials and methods'' for details). This change improved the match of the Gessner \textit{et al}.\ method and the naive rejection sampling method for ST-MNDs for dimension $n=7$ (plots in the right column of Fig~\ref{fig:results_dimensions_6_and_7}), suggesting that accounting for correlation of samples may be an important issue to consider when estimating quantities in higher-dimensional ST-MNDs.

\begin{figure}[!h]
\includegraphics[width=\textwidth]{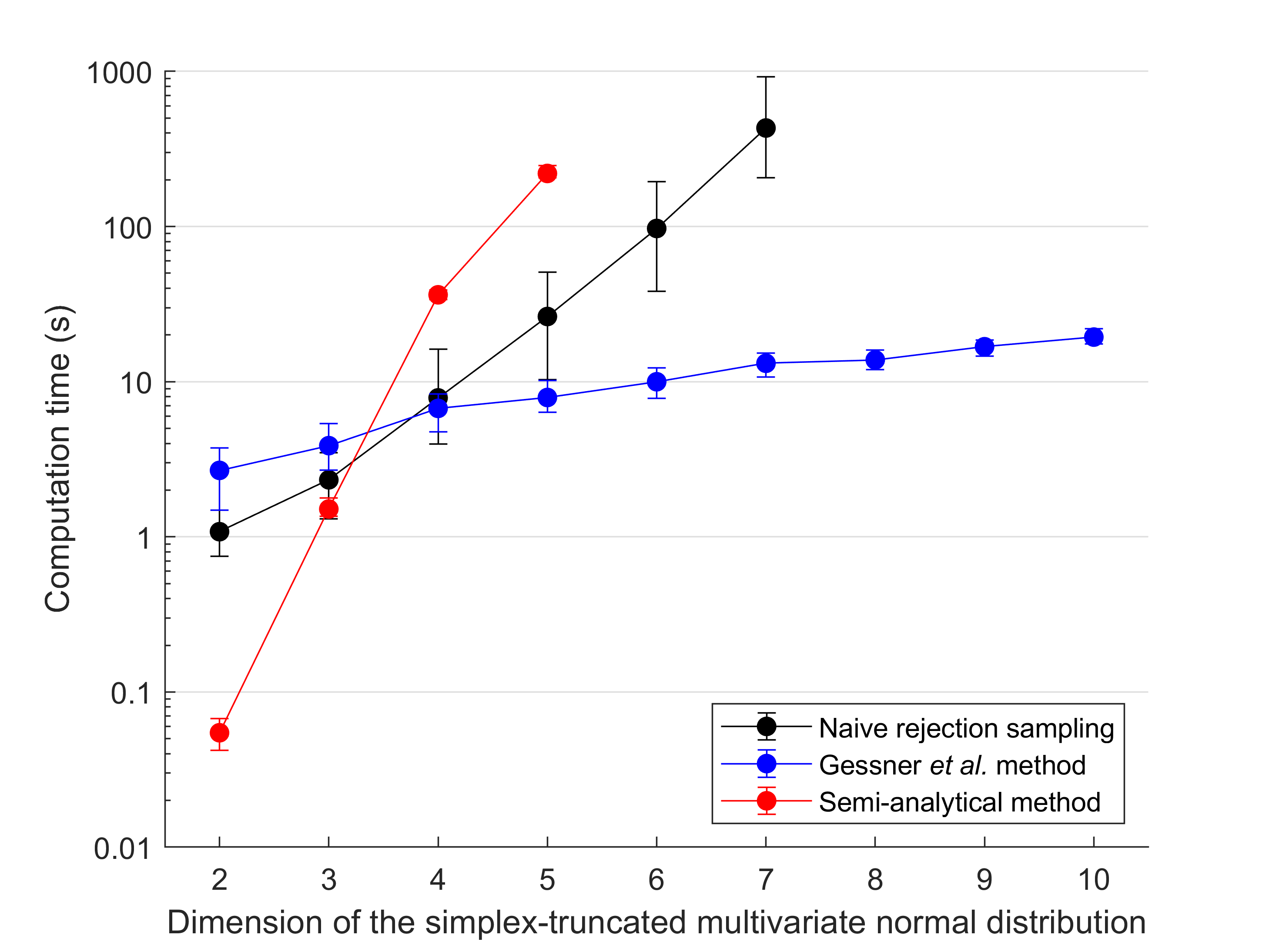}
\caption{{\bf Comparison of computation times (dots and error bars indicating the median and 68\% central credible interval of 100 different values, respectively), for the present implementation of the three methods for calculating the integral, mean and covariance of simplex-truncated multivariate normal distributions.} Notice that the semi-analytical method was the fastest for distributions of low dimension $n \leq 3$. As the dimension of the distribution increases, the Gessner \textit{et al}.\ method becomes increasingly recommended because of its high efficiency, the latter of which is due primarily to this method's ability to sample without rejection. For the naive rejection sampling and semi-analytical methods, results are only shown for distributions of dimension $n \leq 5$ and $n \leq 7$ respectively, as the computational times of both of these methods are excessive at higher dimensions.}
\label{fig:results_computation_time}
\end{figure}

Estimates of the integral, mean or covariance for ST-MNDs of dimension $n > 5$ were not calculated from the semi-analytical method because even without improving its accuracy, this method was far more computationally expensive than the two sampling methods at $n=5$ (Fig~\ref{fig:results_computation_time}). (When the accuracy of the semi-analytical method was improved, as seen in the change of predictions from the central to the right panels shown in Fig~\ref{fig:results_dimension_5}, the median computational cost increased by approximately 10-fold; results not shown.) Similarly, estimates of the integral, mean or covariance for ST-MNDs of dimension $n > 7$ were not calculated from naive rejection sampling because of its substantial computational expense at $n=7$ (Fig~\ref{fig:results_computation_time}).

When comparing the speed of the three methods for ST-MNDs ranging from dimensions $n=2$ to $n=10$ it was found that, in the present implementation, the semi-analytical method was the fastest of the three methods tested when $n \leq 3$, and the Gessner \textit{et al.}\ method was the fastest of the three methods tested when $n \geq 4$ (Fig~\ref{fig:results_computation_time}). The differences in computation speed between the methods are quite substantial; for example, to calculate the desired quantities (integral, mean and covariance) for the last of the 100 ST-MNDs tested with dimension $n=2$ (whose corresponding non-truncated MND possesses mean elements $\mu_1 \approx 0.45$, $\mu_2 \approx 0.28$, and covariance matrix elements $\sigma_{11} \approx 0.17$, $\sigma_{22} \approx 0.06$, $\sigma_{12}=\sigma_{21} \approx 0.04$), our implementations of the semi-analytical method, naive rejection sampling method, and Gessner \textit{et al}.\ method computed all of these quantities in 0.08~seconds, 1.1~seconds and 2.2~seconds respectively.

However, the exact dimension at which one method is more efficient than the other will depend strongly on the details of the implementation (e.g.\ programming language, number of samples and thinning ratio used, algorithm used to estimate integrals of hyperrectangularly-truncated MNDs, etc.). For example, when the thinning ratio of the algorithms used within the Gessner \textit{et al}.\ method to unbiasedly estimate the integral, mean and covariance of ST-MNDs was increased from two to ten (central and right columns of Fig~\ref{fig:results_dimensions_6_and_7}), it was unsurprisingly seen that the median computational time of this method increased approximately by a factor of five (results not shown). However, if Fig~\ref{fig:results_computation_time} had instead showed results for the Gessner \textit{et al}.\ method possessing a five-fold higher computation time, it would still be concluded that this method is the most recommended for ST-MNDs of higher dimension (although the naive rejection sampling method may instead have been favoured at $n=4$). Given these complexities associated with implementation details, what the results \textit{generally} suggest is that the semi-analytical method may be advantageous for fast computations on ST-MNDs of low dimension, but the method of Gessner \textit{et al.}\ may be the only practical method (in terms of computational cost) of those tested here if calculations for higher-dimensional ST-MNDs are sought.

Whilst the Gessner \textit{et al}.\ and semi-analytical methods have potential advantages in computational speed (depending on the dimension of the ST-MND), they are both more onerous to code than naive rejection sampling. The ease of coding naive rejection sampling, combined with its relatively low computational costs for ST-MNDs of low dimension, may make it an acceptable option for certain applications. Increasing accuracy of the naive rejection sampling method can also simply proceed by increasing the number of samples, whilst the other two methods have additional settings that require tuning. In the Gessner \textit{et al}. method, one must additionally choose the thinning ratios for each algorithm, and the conditional probability target value and (relatively small) number of samples used in subset simulation. For the semi-analytical method, one must carefully choose the function that is used to estimate the integral of hyperrectangularly-truncated MNDs, as well as the accuracy settings used in this function.

It is clear, therefore, that all three methods may find practical usage as the ``best'' method in different applications. As a synthesis of the present work, Table \ref{table:methods_summary} provides a overview of the advantages and disadvantages of each of the three methods presented in this manuscript for computations on ST-MNDs.

\begin{table}[!ht]
\centering
\caption{{\bf Summary of the advantages and disadvantages of the three methods presented in this manuscript, for calculations on the simplex-truncated multivariate normal distribution.}}
\begin{tabular}{lccc} 
\\ \hline
				  & Naive rejection & Gessner \textit{et al}.\ & Semi-analytical \\
Method properties & sampling & method & method \\
\hline
Obtain samples     & \checkmark & \checkmark & \text{\sffamily X} \\
Calculate integral & \checkmark & \checkmark & \checkmark \\
Calculate moments  & \checkmark & \checkmark & \checkmark \\
$\quad$ (e.g.\ mean, covariance) \\
Ease of coding and tuning   & High & Medium & Medium \\
Speed of accurate computation for: \\
$\quad$ Low dimensional distributions & Fast & Medium & Very fast \\
$\quad$ High dimensional distributions & Very slow & Medium & Very slow \\
\hline
\end{tabular}
\label{table:methods_summary}
\end{table}

\section*{Discussion}

The present work has two primary contributions. Firstly, to the author's best knowledge, this work provides the first comparison of methods for estimating the integral, mean and covariance of MNDs truncated to the non-negative space under a unit simplex (i.e.\ ST-MNDs). Secondly, one of the tested methods (the semi-analytical method) is a new method which is a novel combination of the inclusion-exclusion principle~\cite{Edelsbrunner1995} and the moment generating function approach for hyperrectangularly-truncated MNDs~\cite{Manjunath2021}, for estimation of these quantities. The calculations that this manuscript details, and the guidance provided around their best usage (Table~\ref{table:methods_summary}), have immediate application for modelling of compositional data which, in the absence of the truncations associated with this data type, would otherwise be normally distributed.

\subsection*{Computation times of all methods depend greatly on dimension}
The semi-analytical method was found to be the fastest for calculating the integral, mean and covariance for ST-MNDs of low dimension (Fig~\ref{fig:results_computation_time}); in the present computational implementation it was the fastest method for dimensions $n=2$ and $n=3$. However, for higher dimensional ST-MNDs, the computational cost of this method increased rapidly; this is not surprising given that Eq~\eqref{eq:Z_general}-\eqref{eq:mu_ij_general} suggest that each extra dimension drastically increases the number of integrals of hyperrectangularly-truncated MNDs that require estimation. In contrast, it was found that the Gessner \textit{et al}.\ method was the only practical method of the three tested here for calculating the integral, mean and covariance of higher-dimensional ST-MNDs. This can be attributed to the ability of the analytical version of elliptical slice sampling, utilised within the Gessner \textit{et al}.\ method, to sample from the domain of the ST-MND with 100\% acceptance rate, regardless of dimension.

Independent of implementation details, each of the three methods has different theoretical computational complexities that can be briefly summarised as follows. The naive rejection sampling method is very straightforward to implement as it involves sampling from MNDs and rejecting samples that fall outside of the required simplex domain, so this method will become computationally impractical when the fraction of samples rejected is large. The semi-analytical method requires computation of various quantities, of which the number (and dimension) of integrals of hyperrectangularly- truncated MNDs is the primary limitation on this method's speed and accuracy. Finally, the Gessner \textit{et al.} method involves a Markov chain Monte Carlo algorithm which samples with 100\% acceptance rate, and is theoretically limited primarily by (1) the additional calculations required to transform samples from a uniform one-dimensional distribution to its equivalent location within the $n$-dimensional simplex region and (2) obtaining a sufficient number of samples to minimise the effects of sample autocorrelation. The effects of these theoretical limitations on computational practicality differ primarily with the dimension of the ST-MND, as was seen in the results from our implementation of the three tested methods.

\subsection*{Calculating other relevant quantities of the simplex-truncated multivariate normal distribution}
Beyond calculation of the integral, mean and covariance of a ST-MND, the methods described here also permit several other useful quantities to be estimated. For example, the integral of a ST-MND can be used to convert relative probability densities to absolute probability densities within the simplex domain. Furthermore, if calculations for an $n$-dimensional ST-MND are being applied to normally-distributed compositional data possessing non-redundant fractions $x_1,...x_n$, the mean and covariance matrix elements associated with the redundant fraction $1 - \displaystyle\sum_{i=1}^n x_n$ can be obtained from calculations of the expectations of this redundant fraction, the square of the redundant fraction, and the product of the redundant fraction with other fractions. These expectations are easily obtained from the sample-based methods. For the semi-analytical method, these expectations require some additional calculations beyond those presented in this manuscript. However, these expectations will ultimately be expressable in terms of first and second moments of hyperrectangularly-truncated MNDs with zero mean (for which general expressions are provided in Eq~\eqref{eq:E_Wi} with its upper limit of summation changed from 2 to $n$, and Eq~\eqref{eq:E_WiWj_general}). Thus, the present research also provides useful methodological support for other computations on ST-MNDs and their applications to compositional data.

\subsection*{Alternative methods for calculations on the simplex-truncated multivariate normal distribution}
Although the three methods described here are particularly useful if simultaneous computation of the integral, mean and covariance of ST-MNDs is sought, there are other methods available which are suitable for calculating a subset of these three quantities, or potentially all of these quantities if further methodological unifications are introduced. For example, Koyama~\cite{Koyama2020} recently detailed how to use the holonomic gradient method to estimate the integral of a MND truncated within a simplex region. Several methods to efficiently sample from the simplex are available, as reviewed in \cite{Altmann2014}, and additional relevant sampling methods have been introduced since that review (e.g.~\cite{Pakman2014,Cong2017,Chaudhry2021}). For example, Cong \textit{et al}.\ \cite{Cong2017} provides an algorithm which can be applied to ensure that samples proposed by naive rejection sampling definitely have all fractions (including the redundant fraction) sum to one.

Sampling methods that attempt to reduce computations outside of the simplex domain, can potentially estimate the mean and covariance of ST-MNDs more efficiently than the naive rejection sampling method. Since sampling from a distribution is a less computationally challenging problem than obtaining the integral of a distribution (and in the present work we are interested in methods that can estimate the integral as well as the mean and covariance of a ST-MND), such efficient sampling methods can often scale well to dimensions much higher than those investigated in the present work. However, alternative sampling methods may also be used to efficiently estimate the integral of ST-MNDs if they are appropriately combined with suitable algorithms for estimating rare event probabilities~\cite{Kroese2011}. For example, alternative sampling algorithms could be used within the Gessner \textit{et al}.\ method to replace the LIN-ESS algorithm used within that method. Since the Gessner \textit{et al}.\ method is particularly well-suited for computations in higher-dimensional ST-MNDs, the availability of multiple alternative sampling options for potential usage within this method suggests the possibility that further accuracy and/or efficiency gains may be possible for computations on higher-dimensional ST-MNDs beyond those seen here.

\subsection*{Future usage of the semi-analytical method}
Specific results regarding computation accuracy (Fig~\ref{fig:results_dimension_2}-\ref{fig:results_dimensions_6_and_7}) and computation time (Fig~\ref{fig:results_computation_time}) for the three methods will be unavoidably dependent on the implementation that is used. For example, dimension-specific changes in the accuracy of the semi-analytical method (compare plots in the right columns of Fig~\ref{fig:results_dimension_2}-\ref{fig:results_dimension_4}, and central and right columns of Fig~\ref{fig:results_dimension_5}) depended strongly on the details of the algorithms for estimating integrals of hyperrectangularly-truncated MNDs that are built into the mathematical software package used here (MATLAB R2021). In other software packages, as well as in future versions of the software package used here, these algorithms are likely to differ in speed and accuracy. Hence, the broad and continuing usefulness of the semi-analytical method will strongly depend on how the programming of modern mathematical and statistical packages, especially the algorithms they possess for computing integrals of hyperrectangularly-truncated MNDs, are improved in the future.

\subsection*{Guidance for calculating only the integral, mean or covariance}
In the present work, the computational costs of calculating individual quantities characterising ST-MNDs were not separately compared. (For example, the methods were not compared for their ability to accurately and efficiently compute \textit{only} the integral). However, some general guidance can be provided here. The computational cost of the naive rejection sampling procedure is the same regardless of whether calculation of the integral, mean and/or covariance are sought, because the same number of samples need to be drawn regardless of which of these quantities is being computed. In the Gessner \textit{et al}.\ method, computational costs are reduced if only estimation of the mean and/or covariance are sought, or if only estimation of the integral is sought. (The computational cost of estimating the mean, covariance, or both, is the same in the Gessner \textit{et al}.\ method.) Furthermore, estimating the mean and/or covariance will draw a lower computational cost compared to the integral, since obtaining a sample mean and covariance from the Gessner \textit{et al}.\ method only requires elliptical slice sampling without any other algorithms. In contrast, estimation of the integral using the Gessner \textit{et al}.\ method requires elliptical slice sampling in tandem with subset simulation and the Holmes-Diaconis-Ross algorithm. In the semi-analytical method, the computational cost of estimating the integral is smaller than the computational cost of estimating the mean, and this latter cost is smaller than the computational cost of estimating the covariance matrix.

\subsection*{Extensions to other convex polytopes and higher order moments}
Although the main focus here was on MNDs truncated in the non-negative space under a unit simplex due to this distribution's applicability to compositional data, all three tested methods are easily extendable to finding the integral, mean and covariance of an MND truncated within \textit{any} convex polytope, as follows. In the naive rejection sampling method, the sample acceptance criterion simply changes to the constraints defining the convex polytope. In the Gessner \textit{et al}.\ method, which requires the linear domain constraints bounding the space $\bm{y}$ to be defined by a matrix $A$ and vector $\bm{c}$ according to the vector inequality given in Eq~\eqref{eq:constraints}, appropriately chosen definitions of $A$ and $\bm{c}$ that differ from their forms specified for ST-MNDs in Eq~\eqref{eq:constraint_A} and Eq~\eqref{eq:constraint_c} would be all that is required. Similarly, for the semi-analytical method, which requires the half-spaces of $\bm{x}$ excised from the allowable domain to be characterised by constraints of the form $\bm{c}_{\bm{v}} < T_{\bm{v}} \bm{x} < \bm{d}_{\bm{v}}$ (e.g.\ Table \ref{table:R}), the associated matrices $T_{\bm{v}}$, and vectors $\bm{c}_{\bm{v}}$ and $\bm{d}_{\bm{v}}$, would require appropriately chosen definitions to suit the half-spaces excised to construct the convex polytope. Additionally, the inclusion-exclusion identities required by the semi-analytical method for calculating the integral, mean and covariance of a convex polytope may differ from those shown in Eq~\eqref{eq:Z_general}-\eqref{eq:mu_ij_general} only by the number of terms present in these equations (as the number of terms depends on the number of linear domain constraints defining the convex polytope rather than its dimension).

The methods presented here can also be generalised to find higher order moments of the ST-MND (or any convex polytope for that matter). For the sampling methods, this generalisation trivially requires calculating different sample expectations from the accepted samples. For the semi-analytical method, new formulae which combine the inclusion-exclusion principle \cite{Edelsbrunner1995} with the moment generating function approach for hyperrectangularly-truncated MNDs \cite{Manjunath2021} are required; it is expected that the derivations of these formulae will be very similar to those obtained for the mean and covariance described here.

\section*{Acknowledgments}
Chris Drovandi, Karlo Hock and Brodie A. J. Lawson are thanked for helpful discussions during manuscript development. This work was funded by an ARC Discovery Early Career Researcher Award DE200100683.

\bibliography{Bibliography}

\clearpage

\section*{Supplementary Material S1: Identifying all ellipse arcs within a linearly truncated domain for the LIN-ESS algorithm}

\setcounter{equation}{0}
\renewcommand{\theequation}{S1.\arabic{equation}}

In the ``Materials and methods'' section of the manuscript it was stated that the third step of the LIN-ESS algorithm for ST-MNDs is to identify all values of $\theta$ on the ellipse defined by Eq~(9) that fall within the domain of the ST-MND. Here the mathematical details required to accomplish this third step of the LIN-ESS algorithm are provided, for any domain truncated by linear constraints (of which ST-MNDs are a special case). The mathematical procedure described here is similar to that provided in Gessner \textit{et al}. (reference [13] in the manuscript text), although some different trigonometric relations are utilised, so these relations are described carefully here.

Finding all values of $\theta$ within the truncated domain requires finding all values of $\theta$ that act as boundaries between arcs of the ellipse falling inside and outside this domain. For example, in the sampling step shown in Fig~1, there are four such boundary values of $\theta$ which form end-points for each of the two ellipse arcs that are within the truncated domain (thick red arcs in Fig~1).

Boundary values of $\theta$ will typically occur at the intersection of the ellipse and one of the domain constraints. (It is possible that these boundary values of $\theta$ could occur at the intersection between the ellipse and more than one domain constraint simultaneously, but this is highly unlikely and thus can be ignored since the vectors $\bm{y}_t$ and $\bm{\nu}$ that generate the ellipse defined by Eq~(9) are both drawn from continuous distributions.) It is denoted that $\theta_i$ represents values of $\theta$ on the ellipse which intersect with the $i$th linear domain constraint. There will typically be zero or two values of $\theta_i$ for each $i$th constraint, because the ellipse will intersect each linear constraint in either zero or two places. (A single value of $\theta_i$ for a constraint indicates a tangential intersection, which is again highly unlikely.) All domain constraints require checking, to identify all values of $\theta_i$.

However, not all intersections $\theta_i$ between the ellipse and a domain constraint will result in a boundary value of $\theta$. For example, in Fig~1, both the $x_1 \geq 0$ and $x_2 \geq 0$ constraints possess intersections with the ellipse that do not separate arcs outside and inside the truncated domain (intersection of red ellipse and dashed black lines in Fig~1). Thus, boundary values of $\theta$ are a subset of the obtained $\theta_i$. However, since the aim is to identify all values of $\theta$ within the truncated domain, it is not necessary to identify which values of $\theta_i$ are boundary values. Instead, one can simply examine all ellipse arcs bounded by adjacent values of $\theta_i$ to see if these arcs fall within the domain or not. Therefore, the main problem then becomes to find all values of $\theta_i$.

Comparing the above considerations with the inequality given by Eq~(6), intersections of the $i$th constraint with the ellipse (and thus the desired values of $\theta_i$) occur when
\begin{equation}
\bm{A}_i \bm{y} + c_i = 0,
\label{eq:intersections}
\end{equation}
where here $\bm{A}_i$ and $c_i$ should be understood as the $i$th rows of $A$ and $\bm{c}$ respectively. (The number of rows in both $A$ and $\bm{c}$ indicates the number of constraints that need to be separately considered for identification of values of $\theta_i$.) From combination of Eq~(9) and Eq~\eqref{eq:intersections}, the problem then becomes the identification of intersection points with a value of $\theta_i$ satisfying
\begin{equation}
\bm{A}_i \left(\bm{y}_t \cos \theta_i + \bm{\nu} \sin \theta_i\right) + c_i = 0,
\label{eq:ESS_angle_equation}
\end{equation}
for each of the constraints. Since the ellipse will typically intersect each linear constraint in either zero or two places, Eq~\eqref{eq:ESS_angle_equation} similarly is expected to have either zero or two solutions for $\theta_i$.

To solve Eq~\eqref{eq:ESS_angle_equation} for $\theta_i$, the non-negative distance $r_i$ is introduced,
\begin{equation}
\label{eq:r_i}
r_i = \sqrt[+]{ \left( \bm{A}_i \bm{y}_t \right)^2 + \left( \bm{A}_i \bm{\nu} \right)^2 }.
\end{equation}
If $|c_i| > r_i$, there are no real solutions for $\theta_i$, which indicates that there are no intersections between the ellipse and the $i$th constraint. In the unlikely event that $|c_i| = r_i$, there is one real solution for $\theta_i$ representing a tangential intersection between the ellipse and the $i$th constraint.

If $|c_i| < r_i$, there are two real solutions for $\theta_i$, indicating two intersections between the ellipse and the $i$th constraint. In this latter case, the angle $\alpha_{i,1} \in (-\pi,\pi]$ is introduced which satisfies
\begin{eqnarray}
\label{eq:alpha_1_condition1} &\cos \alpha_{i,1} = \dfrac{\bm{A}_i \bm{y}_t}{r_i}, \\
\label{eq:alpha_1_condition2} &\sin \alpha_{i,1} = \dfrac{\bm{A}_i \bm{\nu}}{r_i}.
\end{eqnarray}
Substitution of Eq~\eqref{eq:alpha_1_condition1} and Eq~\eqref{eq:alpha_1_condition2} into Eq~\eqref{eq:ESS_angle_equation}, together with usage of an appropriate trigonometric identity, yields
\begin{equation}
\cos \left(\pm \left(\alpha_{i,1}-\theta_i \right)\right) = \dfrac{-c_i}{r_i}.
\label{eq:cosine_alpha1}
\end{equation}
Finally, the angle $\alpha_{i,2} \in [0,\pi]$ is introduced which satisfies
\begin{equation}
\theta_i = \alpha_{i,1} \pm \alpha_{i,2},
\label{eq:theta_i}
\end{equation}
Rearrangement and substitution of Eq~\eqref{eq:theta_i} into Eq~\eqref{eq:cosine_alpha1} yields
\begin{equation}
\cos \alpha_{i,2} = \dfrac{-c_i}{r_i}.
\label{eq:alpha_2_condition}
\end{equation}
Thus, finding the two values of $\theta_i$ indicated by Eq~\eqref{eq:theta_i} requires finding the values of angles $\alpha_{i,1}$ and $\alpha_{i,2}$ which satisfy Eq~\eqref{eq:alpha_1_condition1}, \eqref{eq:alpha_1_condition2} and \eqref{eq:alpha_2_condition}. (As an aside, Eq~\eqref{eq:theta_i} can also be used when there is a tangential intersection, because in this case $\alpha_{i,2}$ will be equal to either zero or $\pi$, thus yielding one value for angle $\theta_i \in [0,2 \pi)$.)

Care must be taken when calculating angles $\alpha_{i,1}$ and $\alpha_{i,2}$ from the inverse trigonometric relations described in Eq~\eqref{eq:alpha_1_condition1}, \eqref{eq:alpha_1_condition2} and \eqref{eq:alpha_2_condition}. It is recommended that each of these angles are calculated in two steps. First, corresponding acute angles, which are denoted here as $\alpha_{i,1}^*$ and $\alpha_{i,2}^*$, both of which are only defined in the first quadrant, $\alpha_{i,1}^*,\alpha_{i,2}^* \in \left[0, \dfrac{\pi}{2}\right]$, can be calculated as
\begin{align}
\label{eq:alpha_1star} \alpha_{i,1}^* &= \arctan \left( \left| \dfrac{ \bm{A}_i \bm{\nu} }{\bm{A}_i \bm{y}_t } \right| \right), \\
\label{eq:alpha_2star} \alpha_{i,2}^* &= \arccos \left( \left| \dfrac{c_i}{r_i} \right| \right).
\end{align}
Second, the values of the required angles $\alpha_{i,1}$ and $\alpha_{i,2}$ are calculated from $\alpha_{i,1}^*$ and $\alpha_{i,2}^*$ via the following formulae:
\begin{align}
\alpha_{i,1} &= \begin{cases} -\pi + \alpha_{i,1}^*, \, & \mbox{if } \bm{A}_i \bm{y}_t < 0 \mbox{ and } \bm{A}_i \bm{\nu} < 0, \\
-\alpha_{i,1}^*, & \mbox{if } \bm{A}_i \bm{y}_t \geq 0 \mbox{ and } \bm{A}_i \bm{\nu} < 0, \\
+\alpha_{i,1}^*, & \mbox{if } \bm{A}_i \bm{y}_t \geq 0 \mbox{ and } \bm{A}_i \bm{\nu} \geq 0, \\
\pi - \alpha_{i,1}^*, & \mbox{if } \bm{A}_i \bm{y}_t < 0 \mbox{ and } \bm{A}_i \bm{\nu} \geq 0,
\end{cases} \label{eq:alpha_1} \\
\label{eq:alpha_2} \alpha_{i,2} &= \begin{cases} \pi - \alpha_{i,2}^*, \quad & \mbox{if } c_i > 0, \\
\alpha_{i,2}^*, & \mbox{if } c_i \leq 0. \end{cases}
\end{align}
In summary, real solutions $\theta_i$ to Eq~\eqref{eq:ESS_angle_equation} are calculated from Eq~\eqref{eq:r_i}, \eqref{eq:theta_i} and \eqref{eq:alpha_1star}-\eqref{eq:alpha_2}, and these solutions exist if and only if $|c_i| \leq r_i$.

Once values of $\theta_i$ have been calculated (if they exist) for all constraints, they are organised in a convenient manner so that arcs of the ellipse that they bound can be systematically checked to see if they are within the truncated domain or not. To accomplish this, found values of $\theta_i$ that are outside of $0 \leq \theta_i < 2 \pi$ are transformed to satisfy this inequality by addition or subtraction of appropriate multiples of $2 \pi$. All found angles $\theta_i$ are then ordered from smallest to largest in a column vector of angles $\bm{\theta}_{\mathrm{F}}$, whose $q$th element is referred to as $(\theta_{\mathrm{F}})_q$. Subsequently, an additional angle, equal to the sum of 2$\pi$ and the smallest $\theta_i$, is appended to the end of this vector, i.e.\ $\bm{\theta}_{\mathrm{F}} \leftarrow [\bm{\theta}_{\mathrm{F}} \,\,\, (\theta_{\mathrm{F}})_1$$+$$2\pi ]^\top$. This additional angle ensures that all parts of the ellipse are bounded by arcs between consecutive elements within $\bm{\theta}_{\mathrm{F}}$. This vector $\bm{\theta}_{\mathrm{F}}$ is a convenient organisation of the found values of $\theta_i$ for arc checking.

All arcs bounded between the angles that are consecutive elements within $\bm{\theta}_{\mathrm{F}}$ are checked to see if the values of $\theta$ these arcs contain fall inside or outside the truncated domain. This checking procedure may be undertaken as follows:
\begin{enumerate}
\item The total number of arcs to be tested, denoted $Q$, is one less than the total number of elements in the vector $\bm{\theta}_{\mathrm{F}}$. Denote $q$ as the current arc to be tested, and set $q=1$.
\item Choose any value of $\theta$ satisfying $(\theta_{\mathrm{F}})_q < \theta < (\theta_{\mathrm{F}})_{q+1}$, and substitute this value of $\theta$ into the ellipse equation, Eq~(9), to identify its location $\bm{y}$.
\item If this value of $\bm{y}$ satisfies all constraints in the inequality given by Eq~(6), then all values of $\theta$ that exist within the arc $(\theta_{\mathrm{F}})_q < \theta < (\theta_{\mathrm{F}})_{q+1}$ correspond to locations within the truncated domain. Otherwise, if any of the constraints in Eq~(6) are not satisfied, the corresponding range of $\theta$ should be disregarded (since these $\theta$ correspond to an arc that falls outside the truncated domain).
\item If the current arc $q$ is equal to total number of arcs $Q$ to be tested, then all arcs have now been checked. Otherwise, set $q \leftarrow q +1$ and go back to Step 2.
\end{enumerate}
This arc checking procedure therefore identifies all values of $\theta$ that exist within the truncated domain, which completes the third step of the LIN-ESS algorithm described in the ``Materials and methods'' section of the manuscript.

\end{document}